\documentclass[fleqn,usenatbib]{mnras}

\usepackage[T1]{fontenc}

\DeclareRobustCommand{\VAN}[3]{#2}
\let\VANthebibliography\thebibliography
\def\thebibliography{\DeclareRobustCommand{\VAN}[3]{##3}\VANthebibliography}

\usepackage{graphicx}	
\usepackage{amsmath}	
\usepackage{amssymb}	

\usepackage{aas_macros}

\title[Constraints on radiative changes after a glitch]{Constraints on wide-band radiative changes after a glitch in PSR~J1452$-$6036}

\author[F.~Jankowski et al.]{
F.~Jankowski,$^{1}$\thanks{E-mail: fabian.jankowski@manchester.ac.uk}
E.~F.~Keane,$^{2,3,1}$
B.~W.~Stappers$^{1}$
\\
$^{1}$Jodrell Bank Centre for Astrophysics, Department of Physics and Astronomy, The University of Manchester, Manchester M13 9PL, UK\\
$^{2}$Centre for Astronomy, School of Physics, National University of Ireland Galway, University Road, Galway, H91 TK33, Ireland\\
$^{3}$SKA Organisation, Jodrell Bank Observatory, Macclesfield, Cheshire SK11 9DL, UK
}

\date{Accepted XXX. Received YYY; in original form ZZZ}

\pubyear{2021}

\begin{document}
\label{firstpage}
\pagerange{\pageref{firstpage}--\pageref{lastpage}}
\maketitle

\begin{abstract}
We present high-sensitivity, wide-band observations (704 to 4032 MHz) of the young to middle-aged radio pulsar J1452$-$6036, taken at multiple epochs before and, serendipitously, shortly after a glitch occurred on 2019 April 27. We obtained the data using the new ultra-wide-bandwidth low-frequency (UWL) receiver at the Parkes radio telescope, and we used Markov Chain Monte Carlo techniques to estimate the glitch parameters robustly. The data from our third observing session began 3~h after the best-fitting glitch epoch, which we constrained to within $\sim 4~\text{min}$. The glitch was of intermediate size, with a fractional change in spin frequency of $270.52(3) \times 10^{-9}$. We measured no significant change in spin-down rate and found no evidence for rapidly-decaying glitch components. We systematically investigated whether the glitch affected any radiative parameters of the pulsar and found that its spectral index, spectral shape, polarisation fractions, and rotation measure stayed constant within the uncertainties across the glitch epoch. However, its pulse-averaged flux density increased significantly by about 10~per cent in the post-glitch epoch and decayed slightly before our fourth observation a day later. We show that the increase was unlikely caused by calibration issues. While we cannot exclude that it was due to refractive interstellar scintillation, it is hard to reconcile with refractive effects. The chance coincidence probability of the flux density increase and the glitch event is low. Finally, we present the evolution of the pulsar's pulse profile across the band. The morphology of its polarimetric pulse profile stayed unaffected to a precision of better than 2~per~cent.
\end{abstract}




\section{Introduction}
\label{sec:Introduction}

Pulsar glitches are sudden spin-up events that happen predominantly in young to middle-aged radio pulsars, as well as in magnetars. With the first one detected in 1969 \citep{1969RadhakrishnanB, 1969Reichley}, 560 glitches are currently listed in the Jodrell Bank glitch catalogue\footnote{\url{http://www.jb.man.ac.uk/pulsar/glitches.html}} \citep{2011Espinoza}, spanning 190 individual pulsars. The known fraction of glitching pulsars is therefore about 7~per cent of the 2871 pulsars that are listed in version 1.64 of the ATNF pulsar catalogue\footnote{\url{https://www.atnf.csiro.au/research/pulsar/psrcat/}} \citep{2005Manchester}, but the overall number might be significantly higher because of observational bias. Initially thought to be the observational results of the cracking of the neutron star's crust in so-called star-quakes \citep{1969Ruderman}, a commonly accepted theory is that they represent catastrophic unpinning events of vortices in the neutron star's superfluid interior \citep{1975Anderson}, that had previously been held in place by the star's strong magnetic field. They are therefore thought to be the observational signatures of angular momentum being released from the superfluid storage located inside the star to the crust. Which part of the superfluid reservoir contributes, i.e.\ the one at the crust-core boundary, or from slightly deeper inside the star, is a matter of open debate (e.g.~\citealt{2012Andersson, 2014Piekarewicz}). Perhaps glitches might even be a combination of the two types mentioned above. A recent review of neutron star physics, including glitches, is given by \citet{2015Haskell}.

Most of our knowledge about glitches comes from the measurements of a small number of pulsars that exhibit either exceptionally high glitch rates, or glitches of the largest observed sizes. The most well-studied glitches are those observed in the Crab (PSR~J0534+2200) and the Vela pulsars (PSR~J0835$-$4510). Those sources have been observed nearly continuously by small professional radio telescopes of the 10 to 30~m class at Jodrell Bank Observatory (e.g.\ \citealt{2011Espinoza, 2015Lyne}), Mt.\ Pleasant \citep{1990Flanagan, 1990McCulloch, 2002Dodson, 2018Palfreyman}, Hartebeesthoek \citep{2010Chukwude}, the 12-m antenna at Parkes \citep{2017Sarkissian, 2019Sarkissian}, and others, and the Vela pulsar due to its high flux density even by amateur telescopes \citep{2019Sarkissian}. Most of the available measurements are from those small telescopes, or from short pulsar timing observations at larger facilities. Despite having excellent cadence, the data are therefore often of low signal-to-noise ratio (S/N), cover a limited bandwidth, or are challenging to calibrate.

The absence of radiative changes after glitches in rotation-powered radio pulsars has historically been an indicator that glitches originate in the neutron star's interior. Most studies so far have found no measureable impact of glitches on the pulsar's magnetospheric configuration or any of the pulsar's radiative parameters. On the other hand, there are a number of examples in the literature where rotation-powered pulsars showed evidence of what are likely magnetospheric changes after glitch events. For example, \citet{2009Lyne} reported a large glitch (and a smaller one) in the Rotating Radio Transient pulsar\footnote{Rotating Radio Transients are believed to be neutron stars with exceptionally high single-pulse variability at radio frequencies, which makes them hard to detect in traditional periodicity searches.} (RRAT) PSR~J1819$-$1458, after which its pulse detection rate and peak pulse energy increased significantly. It also showed a peculiar glitch recovery with an overall \textit{decrease} in spin-down rate, which is in stark contrast to typical glitch recoveries in pulsars. Furthermore, \citet{2011Weltevrede} discovered an unusual double-peaked profile and two additional erratic RRAT-like emission components in the otherwise single-peaked profile of the young, high magnetic field pulsar PSR~J1119$-$6127. The unusual profile behaviour happened shortly after a large-sized glitch in the pulsar, which had a peculiar recovery similar to the one in PSR~J1819$-$1458. Additionally, \citet{2013KeithB} reported a significant increase in correlation between the spin-down rate and the pulse shape parameter of the mode-switching pulsar PSR~J0742$-$2822 after a glitch, which suggested a connection between glitch events and the supposedly magnetospheric mode-switching phenomenon. More recently, \citet{2018Palfreyman} analysed single-pulse data of the 2016 glitch in the Vela pulsar and discovered multiple low-probability events coincident with the glitch epoch: a broad pulse, a missing pulse and two pulses with low linear polarisation, together with changes in the mean and variance of the timing residuals and a dip in flux density.

Analogous to their rotation-powered radio pulsar counterparts, many magnetars have exhibited glitches, and some of them are among the most actively glitching neutron stars known \citep{2014Olausen}. However, glitches in magnetars often show unusual recoveries compared with those in rotation-powered pulsars. They occur sometimes, but not always, simultaneously with radiative changes, which includes flares, pulse profile, spectral hardness, or spin-down torque variations \citep{2003Kaspi, 2008Dib}. The discovery of anti-glitches (sudden spin-\textit{down} events) in a magnetar \citep{2013Archibald} is another peculiar finding.

Here we report on high-sensitivity measurements with the new ultra-wide-bandwidth low-frequency (UWL) receiver at the Parkes 64-m radio telescope, taken serendipitously shortly after a glitch happened in PSR~J1452$-$6036. The data were carefully calibrated, cover a wide frequency range from 704 to 4032 MHz and are of high S/N. We used them to investigate whether any radiative parameters of the pulsar changed significantly after the glitch epoch.

PSR~J1452$-$6036 was discovered in data from the Parkes Multibeam Pulsar Survey (PMPS; \citealt{2001Manchester, 2003Kramer}). It has a catalogued period of about 155~ms, a period derivative of $\sim 1.45 \times 10^{-15} \: \text{s} \: \text{s}^{-1}$ \citep{2003Kramer}, typical for a young to middle-aged pulsar, and a dispersion measure (DM) of $349.54~\text{pc} \: \text{cm}^{-3}$ \citep{2013Petroff}. Its inferred dipolar surface magnetic field strength is $4.8 \times 10^{11}~\text{G}$, it has a characteristic age of $1.69~\text{Myr}$, and a spin-down energy of $1.5 \times 10^{34}~\text{erg}~\text{s}^{-1}$. Only one previous glitch has been reported so far in this pulsar, which occurred on 2009 August 12, and was of small relative size $\Delta \nu / \nu = 29 \times 10^{-9}$ \citep{2013Yu}.

The publication is structured as follows. We summarise our observations in \S~\ref{sec:Observations}, elaborate on our analysis techniques in \S~\ref{sec:Analysis}, present our results regarding any radiative changes post-glitch in \S~\ref{sec:Results}, and discuss them in \S~\ref{sec:Discussion}. Finally, we present our conclusions in \S~\ref{sec:Conclusions}.

\section{Observations}
\label{sec:Observations}

\begin{table}
    \centering
    \caption{Parameters of the observations of PSR~J1452$-$6036 with the Parkes UWL receiver and the Medusa backend reported on in this work. We list the rounded start times as present in the data, the number of frequency channels $n_\text{chan}$ and phase bins $n_\text{bin}$. The observing time was 20~min in all epochs, with 20-s integrations.}
    \label{tab:observations}
    \begin{tabular}{lcccc}
    \hline
    \#      & MJD start         & UTC start$^\text{a}$  & $n_\text{chan}$    & $n_\text{bin}$\\
    \hline
    1       & 58581.460035      & 2019-04-08 11:02:27   & 13312              & 2048\\
    2       & 58582.461887      & 2019-04-09 11:05:07   & 13312              & 2048\\
    3       & 58600.416868      & 2019-04-27 10:00:17   & 6656               & 1024\\
    4       & 58601.400203      & 2019-04-28 09:36:18   & 6656               & 1024\\
    \hline
    \multicolumn{5}{l}{$^\text{a}$ The UTCs in the file names output by the Medusa backend}\\
    \multicolumn{5}{l}{can be offset by one second due to different rounding methods.}\\
    \end{tabular}
\end{table}

We observed PSR~J1452$-$6036 as part of project P1011 ``A wideband survey of pulsars with spectral features'' with the Parkes UWL receiver \citep{2020Hobbs} for 20~min on each of four epochs in 2019~April. The observations were on April~8, 9, 27 and 28, with two epochs either side of the best-fitting glitch epoch. The UWL receiver samples $26 \times 128$-MHz wide native frequency sub-bands that are amplified separately, for a total band of 3328~MHz centred at 2368~MHz. The data were recorded with the GPU-based Medusa backend \citep{2020Hobbs} in pulsar fold mode, with 20-s integrations, 2048 or 1024 phase bins across the profile, 13312 or 6656 frequency channels and were coherently dedispersed within a channel. A reference DM of $349.7~\text{pc} \: \text{cm}^{-3}$ was used for the coherent dedispersion, which is consistent with the pulsar's catalogued value adjusted for its measured secular DM trend \citep{2013Petroff}. Additionally, parts of the band were recorded for cross-validation and redundancy with two other independent backends (CASPSR and DFB4) with 8-s time resolution and similar frequency and phase resolution to the primary data set. We list the parameters of the observations in Table~\ref{tab:observations}.

\section{Analysis}
\label{sec:Analysis}

We calibrated the data in polarisation and absolute flux density using standard techniques with the \textsc{psrchive} software \citep{2004Hotan}, utilising multiple observations of the radio galaxy Hydra~A (3C218) as primary flux density reference. We excised radio frequency interference (RFI) with \textsc{psrchive} tools in the time, frequency and phase-bin domain based on their statistical deviations from the median values or based on the interquartile range of the data. Additionally, a static frequency mask was used to excise channels that were known to be strongly affected by RFI, as well as those at the edges of the 26 native frequency sub-bands, where aliasing effects occur \citep{2020Hobbs}.

\subsection{Pulsar timing analysis}
\label{sec:pulsartiminganalysis}

We first established the pulsar and glitch parameters using standard pulsar timing techniques. For that analysis, we split the Parkes UWL data from each epoch into eight frequency sub-bands of equal width and combined them into eight 2.5-min integrations, which resulted in 256 pulse arrival times (ToAs). We measured ToAs independently for each sub-band to estimate the pulsar's DM together with the rest of the timing parameters. To extend the timing baseline around the glitch so that we could perform a full parameter timing fit, we combined our data set with published ToAs from the pulsar timing programme that is running at the refurbished Molonglo Synthesis Radio Telescope (MOST) as part of the UTMOST project \citep{2017Bailes} at a centre frequency of 835~MHz. \citet{2019Jankowski} presented an overview and the first results from the timing programme. Namely, we used the ToAs published in the second UTMOST pulsar timing paper \citep{2020Lower}. As a preparatory step, we used the \textsc{tempo2} timing software \citep{2006Hobbs} in version 2020.11.1 to determine a value for the phase jump between the Parkes UWL and the UTMOST ToAs, and for an initial timing fit. We then employed the \textsc{pint} pulsar timing software \citep{2021Luo} in version 0.8.1 to simultaneously estimate the timing and glitch parameters in a robust way using Markov Chain Monte Carlo (MCMC) techniques with the \textsc{emcee} ensemble sampler \citep{2013ForemanMackey}. The starting points for the MCMC runs were determined from initial weighted least-squares fits, i.e.\ we started exploring the posterior from close to the maximum likelihood point. For all but the glitch phase increment, we assumed Gaussian priors centred on the best-fitting values from the least-squares fit with standard deviations given by three times the formal parameter uncertainty from the initial fit. For the glitch phase increment, we assumed a uniform prior between $\pm 0.25$ turns that contained all reasonable values, because we could constrain the glitch epoch already somewhat in the initial fit. A wider prior does not change our results. We refined the glitch epoch from the glitch phase increment in the usual way, by iteratively shifting the epoch until the phase increment became approximately zero, and picking the solution that was closest to the original epoch. Standard techniques were used to ascertain that the Markov chains had converged to a high degree. The Jet Propulsion Laboratory Solar System ephemeris DE436\footnote{\url{https://naif.jpl.nasa.gov/pub/naif/JUNO/kernels/spk/de436s.bsp.lbl}} was used to tie our astrometric position to the International Celestial Reference Frame (ICRF). The adopted reference epoch for the period, position and DM measurement is MJD~58320, close to the centre of our combined data set.

In a separate step, we tested whether there were any measurable rapidly-decaying glitch components in the data by adding a single exponentially decaying glitch component to the timing model. We arbitrarily set the exponential glitch decay time to 5~d initially, fit for a decaying step in spin frequency and then explored the parameter space in decay time and decaying step change, together with all parameters above using the same MCMC method. We assumed a uniform prior between zero and 50~d for the decay time, as that approximate range can be constrained with our available timing data, and a Gaussian prior on the decaying step change, initialised from the least-squares fit. We found that the size of the decaying glitch component is not statistically significant, i.e.\ its size is consistent with zero at the $1 \: \sigma$ level and that the posterior distribution of its decay time piles up close to zero. The result is roughly the same if we omit to fit for a step in spin-down rate.

\subsection{Pulse profile analysis}
\label{sec:profileanalysis}

We installed the best-fitting ephemeris as determined from our pulsar timing analysis in the data, and we measured how several radiative pulsar parameters varied over the observing epochs. Our aim was to understand whether any radiative parameters changed significantly after the glitch epoch.

First of all, we visually identified static on-pulse phase gates that contained both profile components, and we measured the band-integrated pulse-averaged flux densities at each epoch. The phase gates were wide enough to include the broader profiles in the lower frequency sub-bands. We show the approximate on-pulse phase ranges below (Fig.~\ref{fig:ProfileEvolution}). We then split the band into 16 frequency sub-bands and fit a simple power law of the form $S(\nu) \propto \nu^{\alpha}$ to the data, where $S$ is the pulse-averaged flux density, $\nu$ is the centre frequency of that sub-band, and $\alpha$ is the spectral index. Our choice of the number of sub-bands ensured an S/N above 15 in each sub-band. As in \S~\ref{sec:pulsartiminganalysis}, we employed MCMC techniques to fit the model to the data, with the start parameters initialised from weighted least-squares fits. Although slight deviations from a simple power law model were apparent in the data, we refer more complex spectral modelling and the necessary model selection process to future work. The derived spectral indices characterised the spectrum sufficiently well. We did not attempt to incorporate flux density measurements from UTMOST data because they currently lack full polarisation information and are not calibrated on an absolute scale.

Secondly, we used the \texttt{rmfit} program that is part of the \textsc{psrchive} software suite to estimate the rotation measure (RM) at each epoch. We performed a brute-force search in trial RM in a range of values around the catalogued RM, $+10(5)~\text{rad}~\text{m}^{-2}$ \citep{2006Han}, typically $\pm 100~\text{rad}~\text{m}^{-2}$, and iteratively decreased the spacing of the grid until the RM range was highly over-sampled. Using a custom program, we fit a Gaussian function to the RM spectrum around the maximum in polarised flux density, and we estimated the parameters robustly as above. We quote the mean of the Gaussian and its uncertainty as best-determined RM. As well as in the Galactic magnetic field that permeates the interstellar medium (ISM), the radio waves from pulsars undergo Faraday rotation in the magnetised plasma of Earth's ionosphere. To estimate the ionospheric Faraday rotation, we used the \textsc{ionfr} software \citep{2013Sotomayor} with maps of the vertical total electron content (TEC) in the ionosphere published by the Crustal Dynamics Data Information System \citep{2010Noll} at NASA. Specifically, we used the TEC maps from the CODE analysis centre\footnote{\url{https://cddis.nasa.gov/Data_and_Derived_Products/GNSS/atmospheric_products.html}} for each day of observations and determined the ionospheric piercing points from the pulsar's coordinates and location of the Parkes telescope. We adopted the ionospheric RM values and their uncertainties from the closest reported hours. As part of \textsc{ionfr}, we employed the coefficients of the 13th version of the International Geomagnetic Reference Field (see~\citealt{2015Thebault}). We report ionospheric-corrected RMs as $\text{RM}_\text{ism} = \text{RM}_\text{obs} - \text{RM}_\text{ion}$, where $\text{RM}_\text{obs}$ and $\text{RM}_\text{ion}$ are the observed and ionospheric RMs, respectively.

Thirdly, we investigated the polarisation fractions of the pulsar's radiation. Namely, we measured the total $P/I$, the linear $L/I$ and the circular polarisation fractions $V/I$ per epoch, where $P = \sqrt{L^2 + V^2}$ is the total polarised flux density, $L$ is the total linearly polarised flux density corrected for noise bias using the method by \citet{2001Everett}, based on the measured quantity $L_\text{m} = \sqrt{Q^2 + U^2}$, and $I, Q, U$ and $V$ are the values of the Stokes parameters.

\section{Results}
\label{sec:Results}

\subsection{Glitch parameters}
\label{sec:GlitchParameters}

\begin{figure*}
  \centering
  \includegraphics[width=\textwidth]{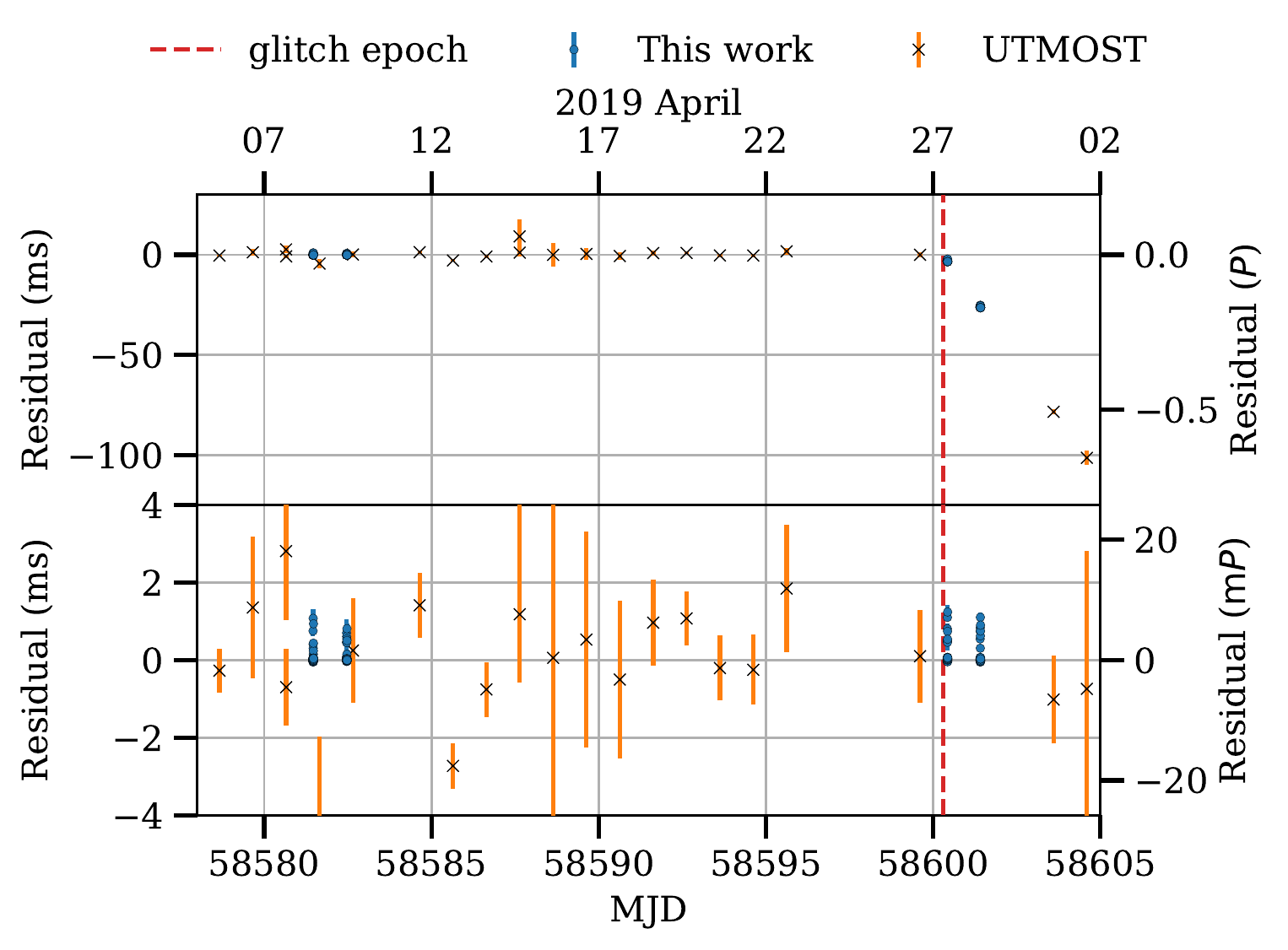}
  \caption{Timing residuals of PSR~J1452$-$6036 zoomed in around the glitch epoch, after fitting for a spin-down and glitch model as described in \S~\ref{sec:GlitchParameters} (bottom panel) and when the glitch term is removed from that model (top panel). In the top panel we display the residuals after the glitch epoch that deviate beyond half a period without wrapping them in phase (see the secondary vertical axis) to show the glitch signature more clearly. It is apparent that our Parkes data dominate the fit near the best-fitting glitch epoch and that the start of our third observing epoch was only 3~h after the event. The slight spread in the Parkes residuals comes from using ToAs from eight frequency sub-bands without accounting for frequency-dependent profile evolution across the wide bandwidth; however, the vast majority of them are centred close to zero.}
  \label{fig:GlitchFit}
\end{figure*}

\begin{table}
    \centering
    \caption{Best-fitting pulsar and glitch parameters from the pulsar timing analysis of PSR~J1452$-$6036. These were derived using MCMC techniques, with the start parameters initialised from weighted least-squares fits. We state the maximum-likelihood values and use the maximum values of the generally asymmetric $1 \: \sigma$ ranges from the posterior distributions as uncertainties. The position is referenced to the ICRF.}
    \label{tab:timingparameters}
    \begin{tabular}{lcc}
    \hline
    Parameter                                       & Value                 & Comment\\
    \hline
    RA (J2000, hms)                                 & 14:52:51.885(4)       &\\
    Dec (J2000, dms)                                & $-$60:36:31.43(5)     &\\
    $t_0$ (MJD)                                     & 58320                 &\\
    $v (t_0)$ (Hz)                                  & 6.45193792529(2)      &\\
    $\dot{\nu}$ ($10^{-14} \text{s}^{-2}$)          & $-$6.0383(2)          &\\
    DM$^\text{a}$ ($\text{pc} \: \text{cm}^{-3}$)   & 349.525(1)            & Timing fit.\\
    $t_\text{g}$ (MJD)                              & 58600.292(3)          &\\
    $\Delta \nu$ ($\mu \text{Hz}$)                  & 1.7454(2)             &\\
    $| \Delta \dot{\nu} |$ ($\text{s}^{-2}$)        & $< 2 \times 10^{-16}$ & $3 \: \sigma$ limit.\\
    $\text{N}_\text{ToA}$                           & 543                   & 256 Parkes.\\
    Data range (MJD)                                & 57955 -- 58690        &\\
    Timescale                                       & TDB                   &\\
    Solar System ephemeris                          & DE436                 &\\
    \hline
    Derived                                         &                       &\\
    \hline
    $\Delta \nu / \nu$ ($10^{-9}$)                  & 270.52(3)             &\\
    $\Delta \dot{\nu} / \dot{\nu}$ ($10^{-3}$)      & < 4                   & $3 \: \sigma$ limit.\\
    \hline
    \multicolumn{3}{l}{$^\text{a}$ Accounts for the profile evolution across the band.}\\
    \end{tabular}
\end{table}

We discovered the glitch in PSR~J1452$-$6036 when we combined our multiple epoch Parkes data. Serendipitously, our observations were spaced equally around the glitch epoch, with the best-fitting glitch epoch, MJD~58600.292(3), being shortly before the start of our third observation on 2019 April 27 (start time MJD~58600.41689). As described in \S~\ref{sec:pulsartiminganalysis}, we robustly estimated the pulsar and glitch parameters using MCMC techniques from our sub-banded Parkes UWL data together with extant UTMOST ToAs. We list the resulting best-fitting values and their uncertainties determined from the posterior distributions in Table~\ref{tab:timingparameters} and we show the post-fit timing residuals around the best-fitting glitch epoch in Fig.~\ref{fig:GlitchFit}. In particular, our data constrain the glitch epoch to within $\sim 4~\text{min}$, and we found it to be only 3~h before the start of our third observing epoch. The glitch is of intermediate relative size with $\Delta \nu / \nu =  270.52(3) \times 10^{-9}$ and is about nine times larger than the only other glitch reported in this pulsar (in 2009 August; \citealt{2013Yu}). We measured no significant change in spin-down rate so that we place a $3 \:\sigma$ upper limit on it, $\Delta \dot{\nu} / \dot{\nu} < 4 \times 10^{-3}$. Our inferred glitch parameters generally agree with those reported by \citet{2020Lower} using UTMOST data alone, but our Parkes UWL data constrain them significantly better.

Additionally, we tested for the presence of a rapidly-decaying glitch component in the data, but found it not to be significant. We conclude that given the sensitivity and coverage of our data, there was only a permanent step in spin frequency at the glitch epoch.

\subsection{Constraints on wide-band radiative changes}
\label{sec:RadiativeChanges}

\begin{figure*}
  \centering
  \includegraphics[width=0.84\textwidth]{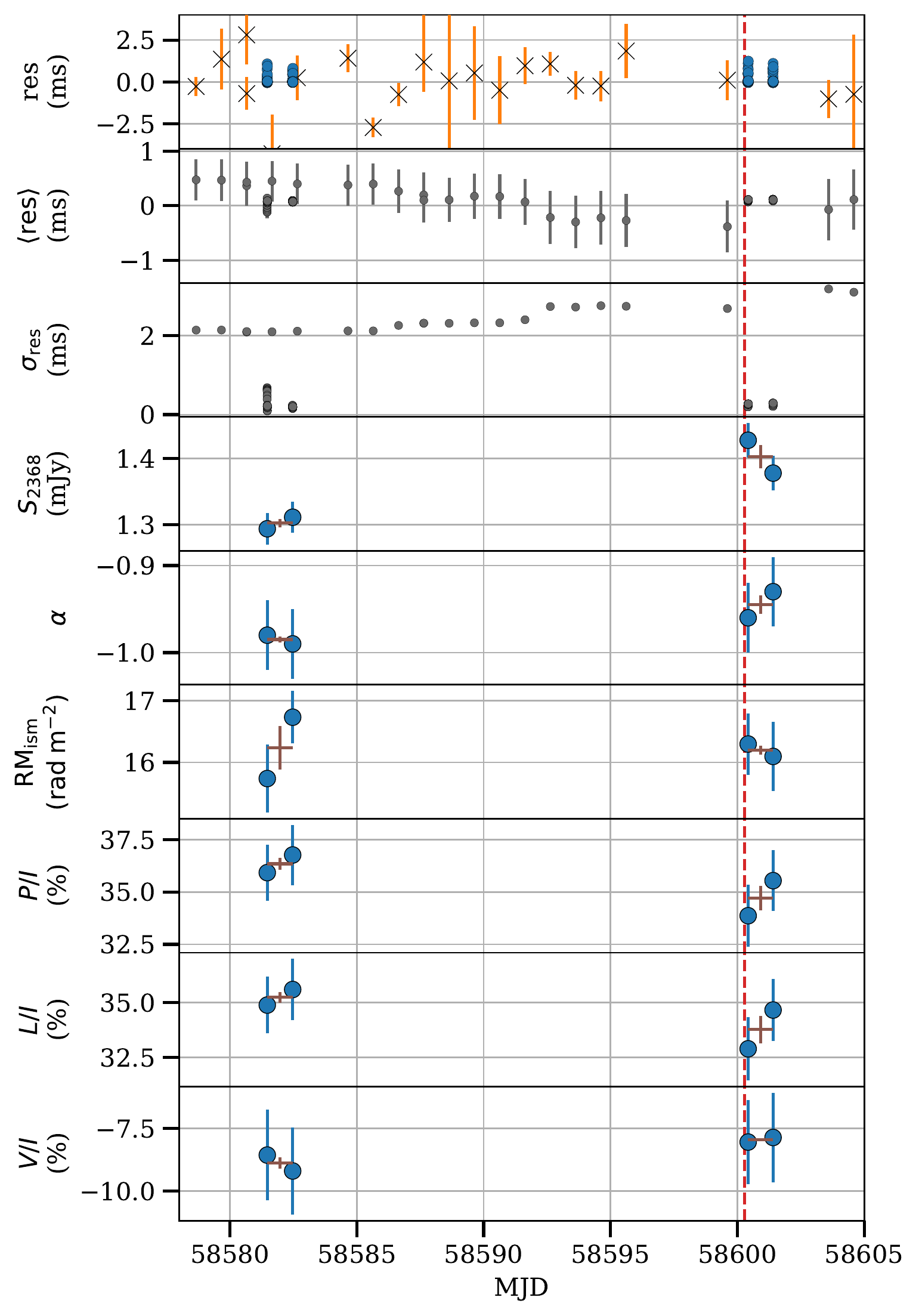}
  \caption{Pulsar parameters plotted against time for PSR~J1452$-$6036. Namely, we show the timing residuals around the glitch epoch, their means and standard deviations computed in sliding boxcar windows of 32 data points, the band-integrated pulse-averaged flux densities for the total intensity, the best-fitting spectral indices assuming a simple power law model, the ionospheric-corrected RMs and the polarisation fractions $P/I$, $L/I$ and $V/I$ at 2368~MHz. We mark the glitch epoch, and we show the mean pre and post-glitch parameter values and their standard errors with brown crosses to help identify significant changes.}
  \label{fig:parametertimeline}
\end{figure*}

\begin{table}
    \centering
    \caption{Mean values of the best-fitting parameters from the profile analysis of PSR~J1452$-$6036. We list the mean values over all epochs and their standard errors at a frequency of 2368~MHz. For the ionospheric rotation measure $\text{RM}_\text{ion}$, we give the mean value and the mean uncertainty.}
    \label{tab:profileparameters}
    \begin{tabular}{lc}
    \hline
    Parameter                                               & Value\\
    \hline
    $S_{2368}$ (mJy)                                        & 1.36(3)\\
    $\text{RM}_\text{obs}$ ($\text{rad} \: \text{m}^{-2}$)  & +15.2(2)\\
    $\text{RM}_\text{ism}$ ($\text{rad} \: \text{m}^{-2}$)  & +16.2(2)\\
    $\text{RM}_\text{ion}$ ($\text{rad} \: \text{m}^{-2}$)  & $-1.06$, 0.14\\
    $\alpha$                                                & $-0.97(1)$\\
    $P/I$ (\%)                                              & 35.5(5)\\
    $L/I$ (\%)                                              & 34.5(5)\\
    $V/I$ (\%)                                              & $-$8.4(3)\\
    \hline
    \end{tabular}
\end{table}

We measured various pulsar parameters that characterise its radiative properties (or those from propagation effects) from either the band-integrated or sub-banded data in each epoch, as described in \S~\ref{sec:profileanalysis}. Namely, we investigated the band-integrated flux density, the spectral index, the DM, the RM, the fractional total, linear and circular polarisations. We show a timeline of these radiative parameters of PSR~J1452$-$6036 together with the timing residuals in Fig.~\ref{fig:parametertimeline}, and we present their mean values in Table~\ref{tab:profileparameters}.

The polarisation fractions, as well as the spectral index and the RM stayed constant within the uncertainties across the glitch epoch. On the contrary, the pulse-averaged flux density showed an increase after the glitch by about 0.12~mJy ($\sim 10$~per cent) with respect to the mean pre-glitch value, and then seemed to decrease to above the pre-glitch level in the next epoch. To estimate the statistical significance of the rise, we computed the mean and its standard error over the pre-glitch values. The flux density increase directly after the glitch (in epoch 3) is somewhat significant at the $4.4 \: \sigma$ level, where $\sigma$ is the combined uncertainty of that flux density measurement and the pre-glitch flux density error. The difference between the mean values of the pre and post-glitch measurements has a significance of $3.5 \: \sigma$, where $\sigma$ is the sum of the standard errors. The flux density increased at all radio frequencies roughly the same, as we discuss in more detail below.

\subsection{Robustness and origin of the flux density change}
\label{sec:originsfluxdensitychange}

To understand whether the increase in flux density is genuine, or merely an artefact from our absolute flux density calibration, we compared the timeline of PSR~J1452$-$6036 with those of four high-DM reference pulsars, PSRs~J1059$-$5742, J1614$-$5048, J1658$-$4958 and J1705$-$3950, that we observed at the same epochs and close in time to the target. We processed their data using the same pipeline. Beneficially, we obtained two observations of PSR~J1614$-$5048 on the day of the glitch event. Aside from PSR~J1059$-$5742 that showed a significant linearly increasing trend, the flux densities of the other three reference pulsars are constant within the uncertainties over the four epochs. Specifically, we did not see any systematic trends for the other reference pulsars. PSR~J1059$-$5742 is special in the sense that with a catalogued DM of 108.7~$\text{pc} \: \text{cm}^{-3}$ \citep{2004HobbsPMPS} it has the lowest DM in our sample of reference pulsars. We, therefore, suspect that its linear trend in flux density is astrophysical in nature (e.g.\ due to scintillation) and unrelated to our calibration procedure, as the remaining reference pulsars show no apparent trends. Importantly, when we compared the relative flux densities of the pulsars normalised by the mean of the pre-glitch measurements, the epoch directly post-glitch (epoch~3) of PSR~J1452$-$6036 stands out as the one with the greatest relative change and the only near 10~per cent in magnitude. Similarly, any astrophysical changes in flux density of our primary calibrator source Hydra A would have imprinted a systematic trend in the data of all reference pulsars in the same way, which is not what we see.

Secondly, the frequency channels excised in the data cleaning process are closely comparable in number and location across the band in all epochs. For this reason, variations in RFI excision cannot explain the flux density difference.

Another critical question concerns whether the rise in flux density is due to scintillation in the ISM. As described above, PSR~J1452$-$6036 has a best-fitting DM of $349.525(1) \: \text{pc} \: \text{cm}^{-3}$, which is substantial and is higher than the DMs of about 80~per~cent of the pulsars in the pulsar catalogue. It is therefore expected to have reasonably stable observed flux densities over time (e.g.~\citealt{2000Stinebring}). Flux density modulation is often characterised using the modulation index
\begin{equation}
    m = \frac{ \sigma_S }{ \left< S \right> },
\end{equation}
where $\sigma_S$ is the standard deviation and $\left< S \right>$ is the mean of the Stokes I flux density time series. Alternatively, those parameters can be substituted by their more robust counterparts, such as the median and the robust standard deviation. For PSR~J1452$-$6036 \citet{2018Jankowski} measured long-term modulation indices of $0.13^{+0.02}_{-0.02}$ and $0.19^{+0.1}_{-0.08}$ between about 1.23 and 1.53~GHz (20~cm) and 2.64 and 3.56~GHz (10~cm), respectively. Naively speaking, the increase in flux density could therefore be entirely due to propagation effects. However, the fractional bandwidth of the UWL receiver is substantially larger than what was used in earlier work, the integration times are slightly longer and the time span over which the increase happened is significantly shorter. Assuming the \textsc{ne2001} Galactic free electron model \citep{2002Cordes}, a transverse velocity of $100~\text{km} \: \text{s}^{-1}$ for the pulsar and the parameters of our observations, such as the effective bandwidth (typically 2.43~GHz) and integration time, we calculated expected modulation indices of about $0.003$, $0.14$ and $0.15$ at 2.368~GHz for strong diffractive and refractive scintillation, and the combined effect. The flux density modulation appears therefore to be dominated by slowly-varying refractive effects, and our theoretical estimates are consistent with the literature measurements stated above. We also confirmed that diffractive scintillation is well quenched. While we cannot exclude that the change is due to refractive scintillation, the timescale over which the increase happened is much smaller than what is usually considered for refractive effects, i.e.\ fading times of multiple months to years with a strong dependence on DM and observing frequency \citep{1982Sieber, 1984Rickett}. Additionally, the size of the increase is close to the maximum magnitude of dispersion for refractive scintillation, both measured and computed. It seems unlikely that it is indeed refractive scintillation that we see, given the short timescale. The slight apparent decrease in flux density in epoch~4 is equally hard to reconcile with it.

It is challenging to quantify the temporal chance coincidence probability between the flux density increase and the glitch event. That is because we do not have enough knowledge of the scintillation parameters of the pulsar in our wide-band observing setup and, more importantly, its glitch size or waiting time distributions are mostly unconstrained. As well as this, there are only two glitches reported in the pulsar in a $\sim 9.7 \: \text{y}$ period, and this is the first large one. If we assume that the amplitude of the flux density modulation due to scintillation is described by a Rice or Rayleigh distribution (e.g.~\citealt{1971Lang}), and we tune its shape, scale and location parameters to match the mean and theoretical estimate for its modulation index (see above), we find that the cumulative probability to measure a value at least as high as in epoch 3 could be up to 25 per cent. Note that we do not have enough measurements to fit the empirical distribution and that our choice of distribution parameters is not unique. This is the probability to measure a flux density amplitude at least as high \textit{independent} of the time between observations, however, what we need is the temporal probability as a function of lag in days (see e.g.~\citealt{2000Stinebring} for a long-term structure function study). The joint probability would then be the product of it with the temporal probability for a glitch of the reported size. Despite not being able to quantify it further, it is, therefore, reasonable to assume that the chance coincidence probability is low.

\subsection{Polarimetric pulse profile}
\label{sec:polarimetricprofile}

\begin{figure*}
  \centering
  \includegraphics[width=0.85\columnwidth]{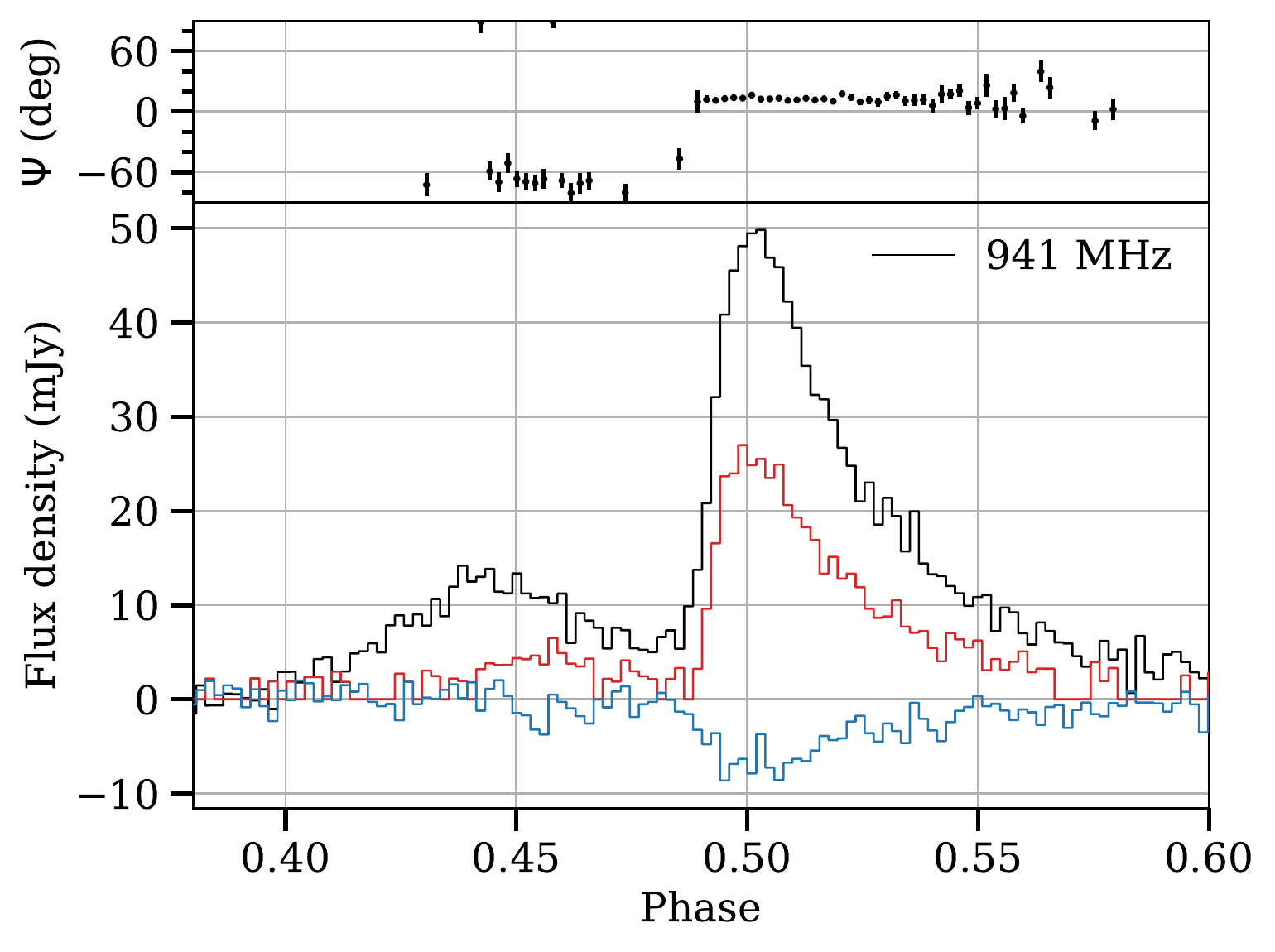}
  \includegraphics[width=0.85\columnwidth]{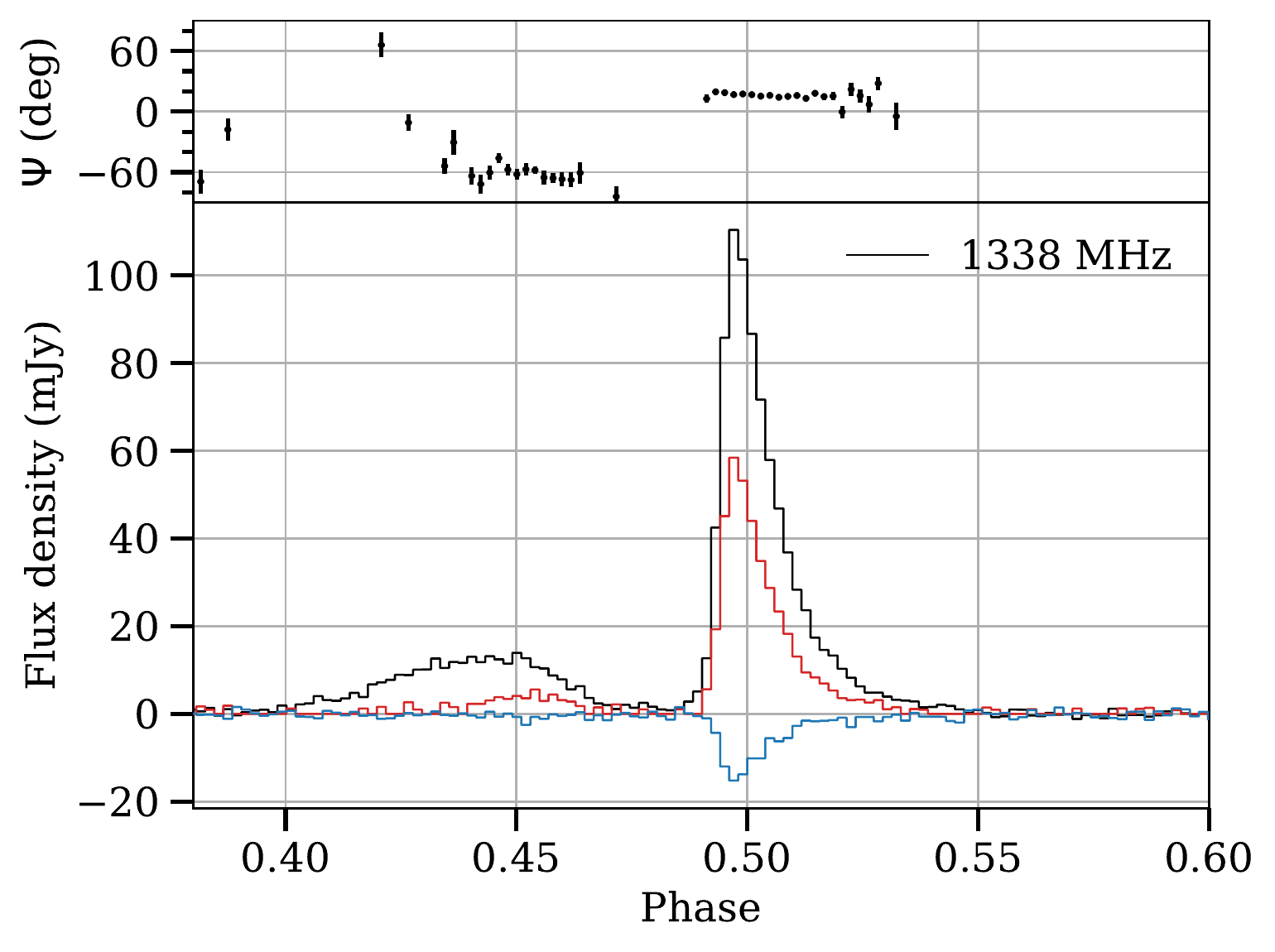}
  \includegraphics[width=0.85\columnwidth]{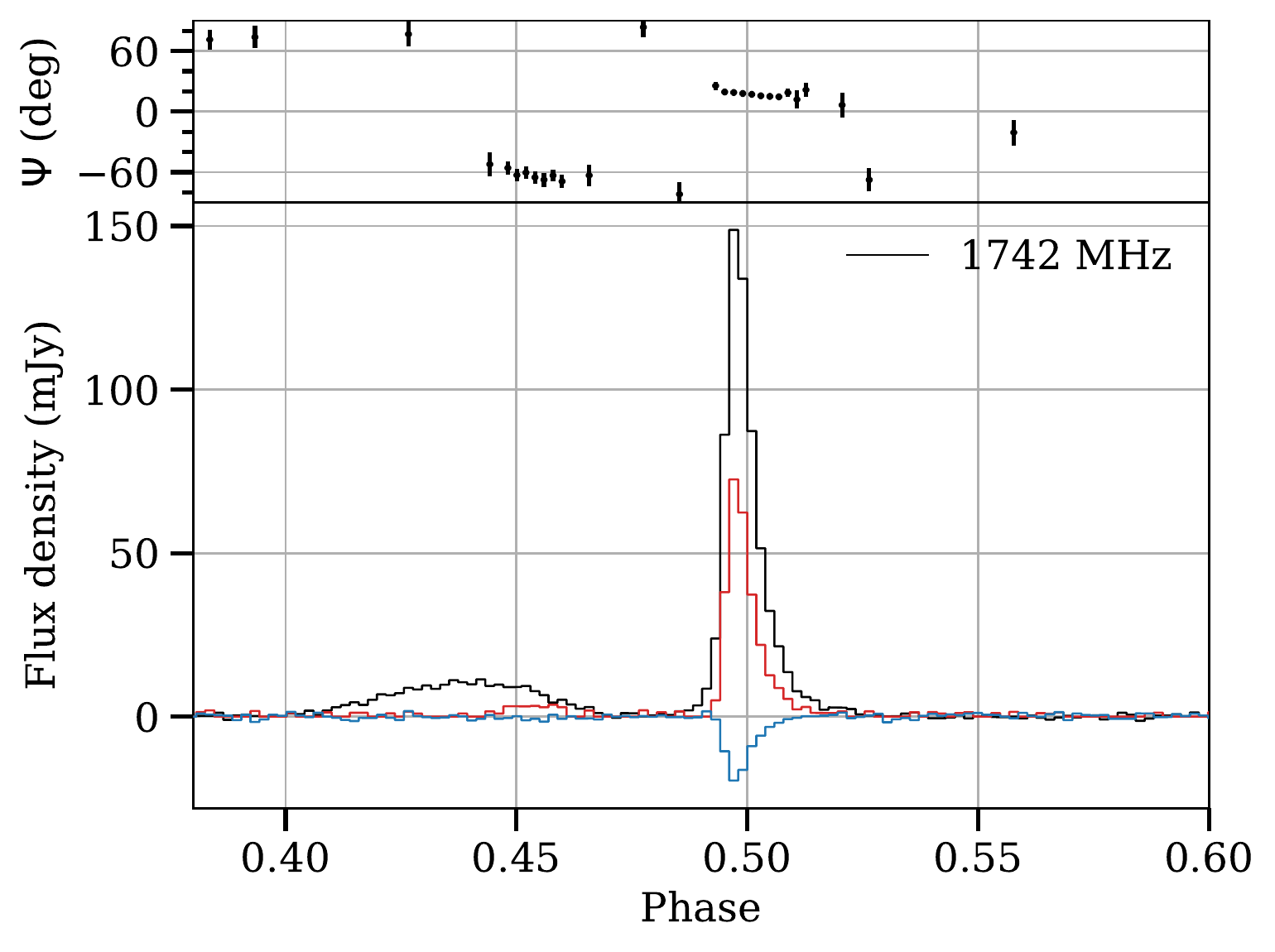}
  \includegraphics[width=0.85\columnwidth]{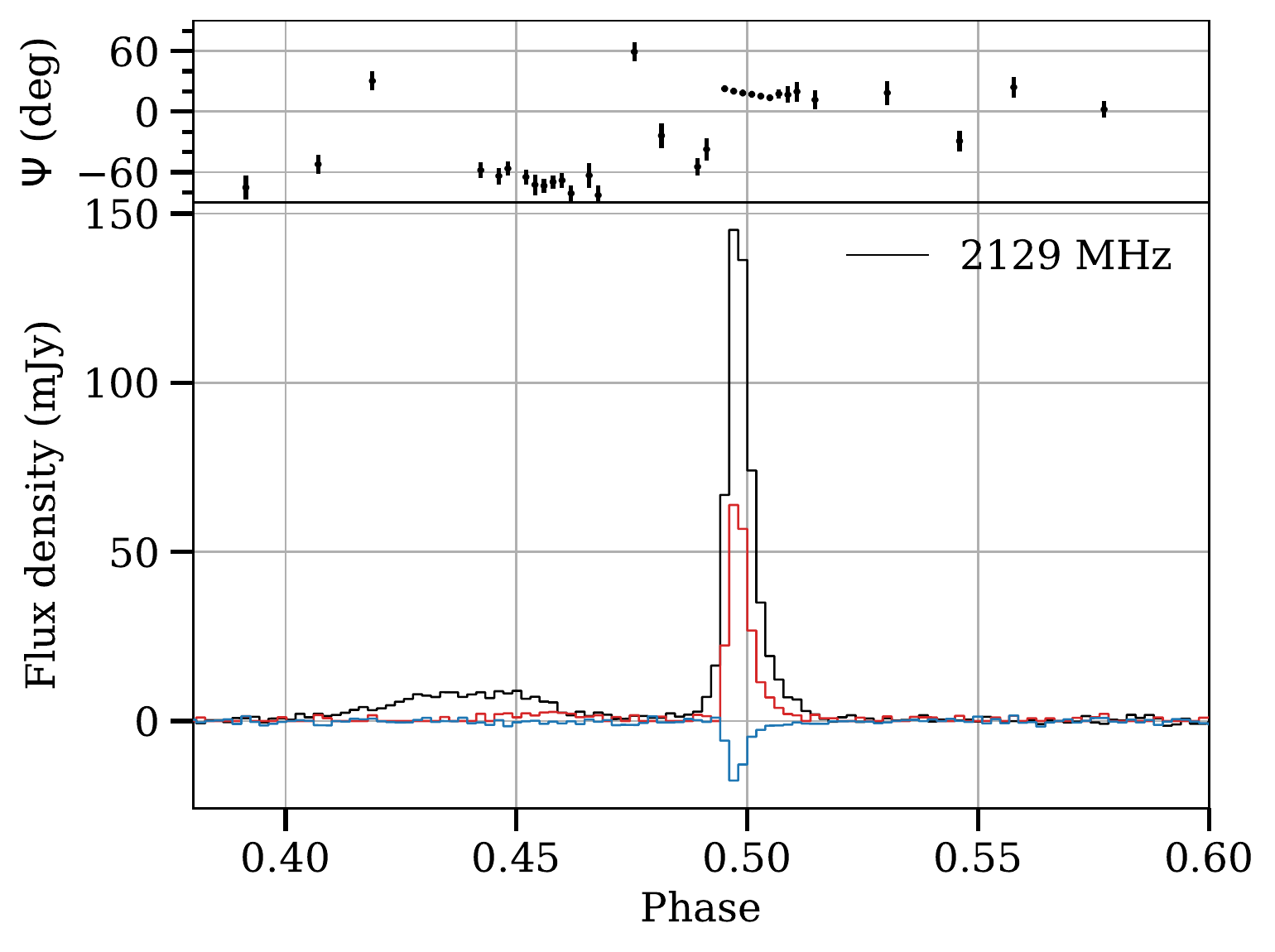}
  \includegraphics[width=0.85\columnwidth]{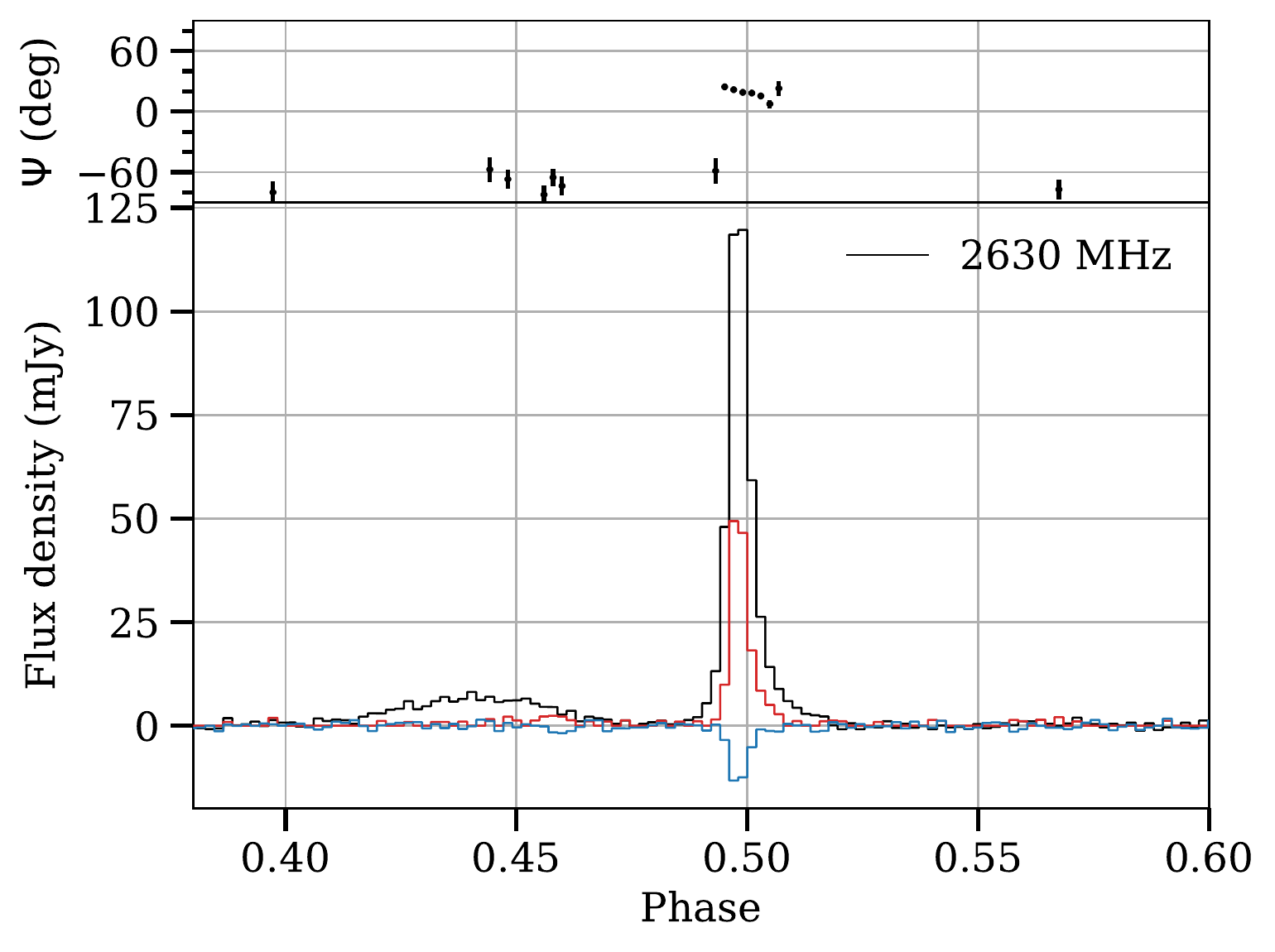}
  \includegraphics[width=0.85\columnwidth]{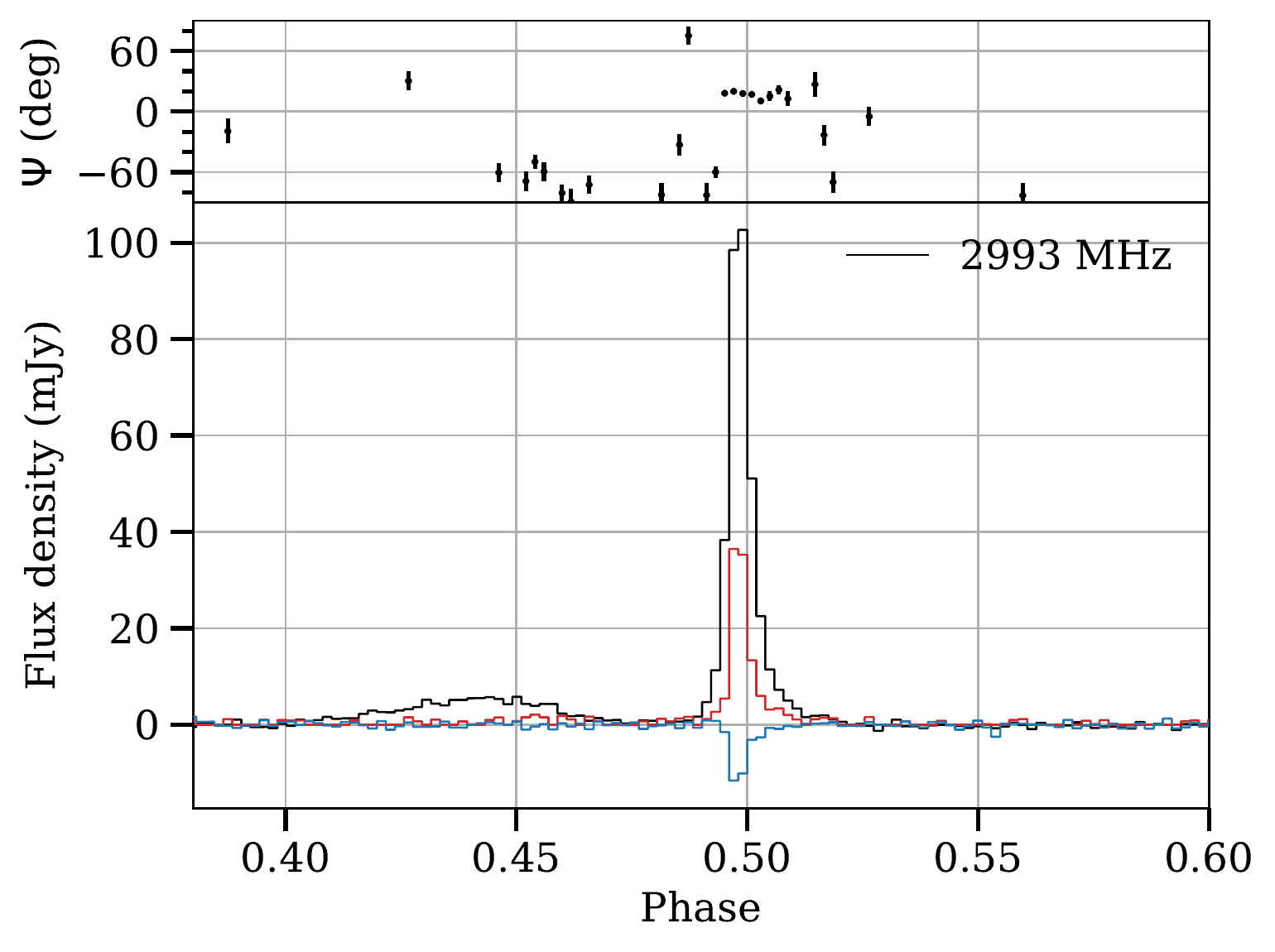}
  \includegraphics[width=0.85\columnwidth]{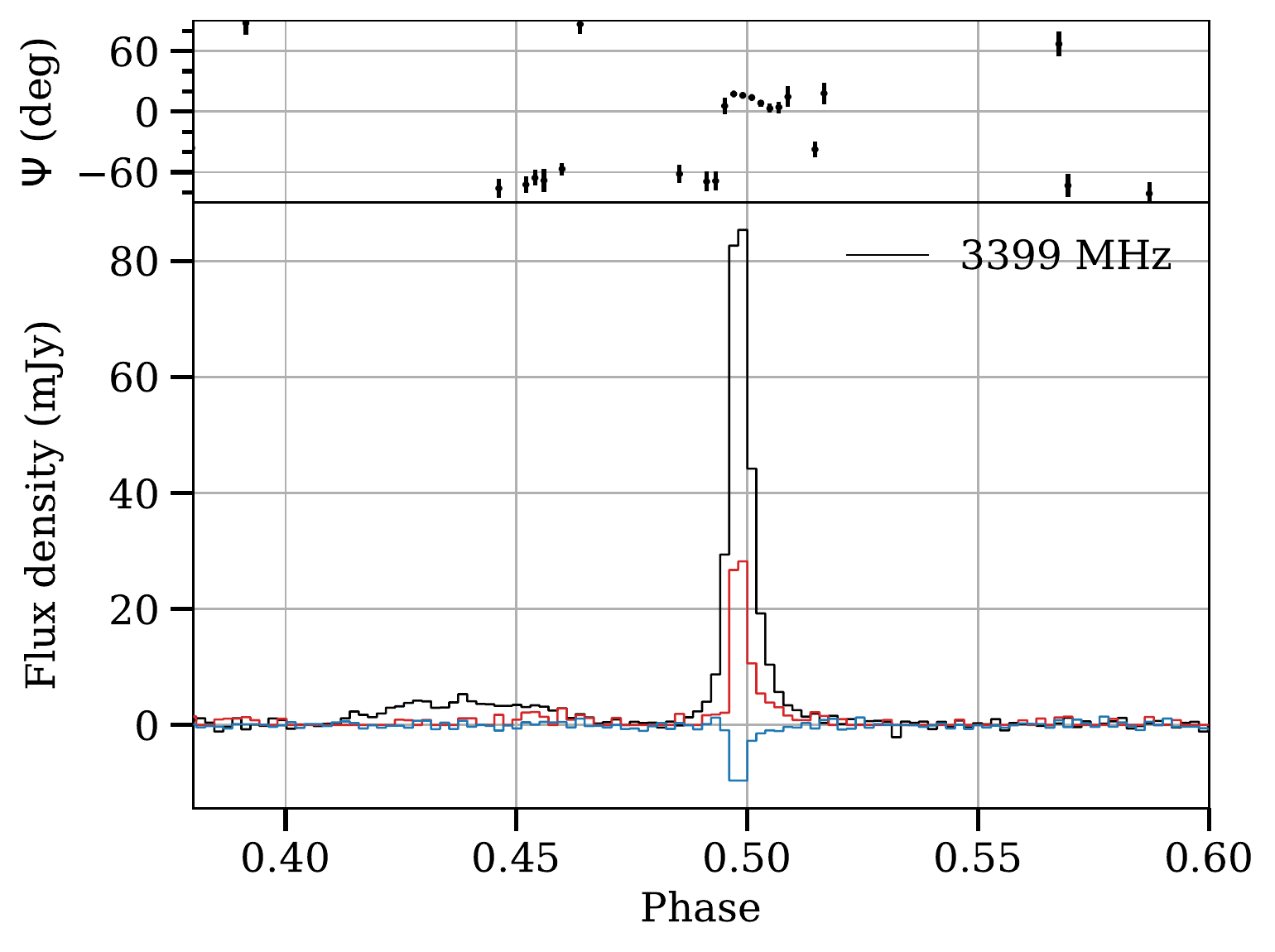}
  \includegraphics[width=0.85\columnwidth]{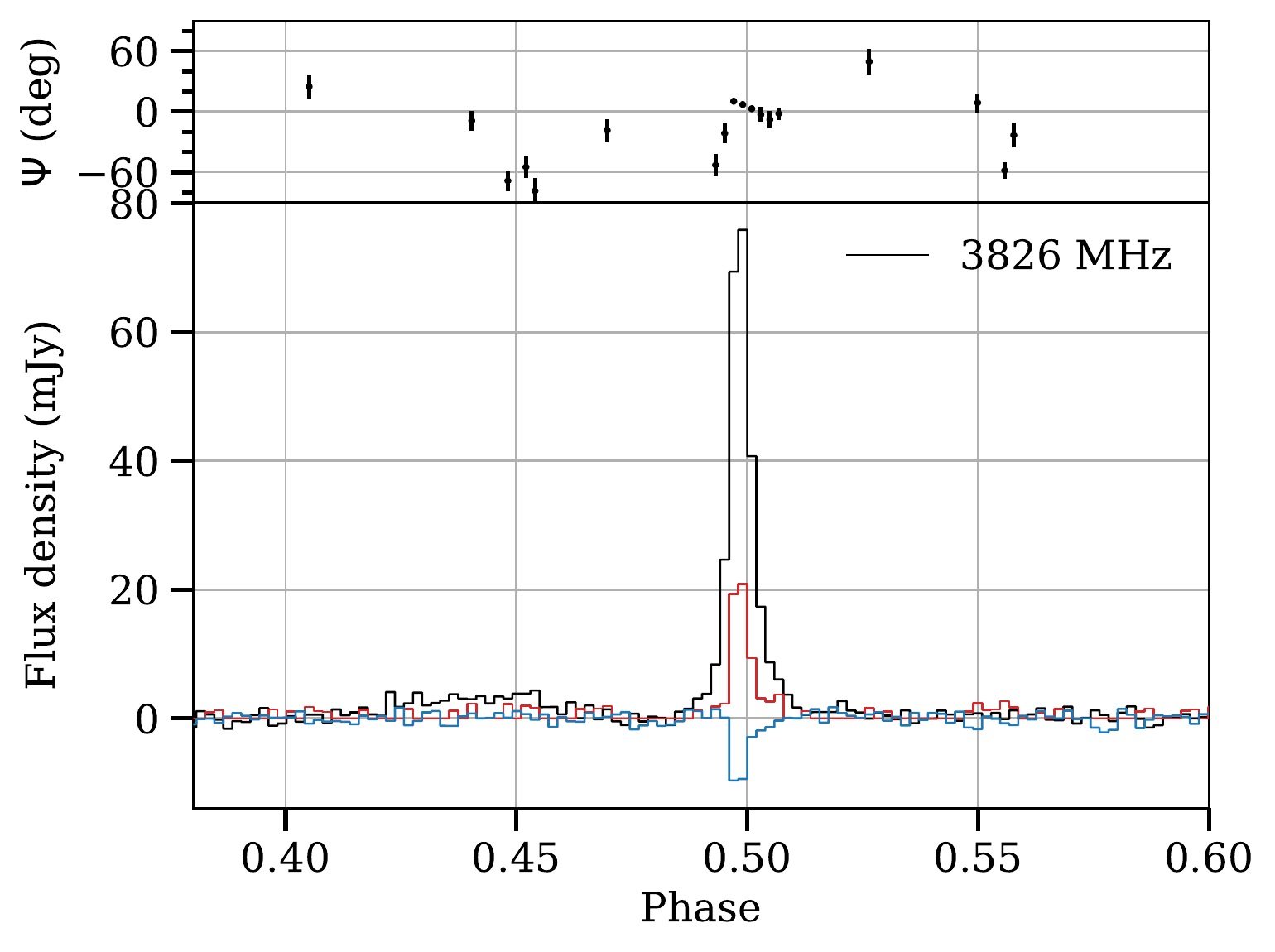}
  \caption{Evolution of the total integrated profile of PSR~J1452$-$6036 with frequency. We show the total intensity (black), the debiased linear (red) and the circular flux densities (blue), as well as the polarisation position angle in eight frequency sub-bands across the 3.3-GHz band of the Parkes UWL receiver in order of increasing centre frequency. The plots show roughly the on-pulse phase range. The position angle is displayed for all phase bins in which the debiased linear flux density exceeds twice the RMS of the Stokes I baseline. Scatter-broadening of the profile due to multi-path propagation in the ISM is visible in the low-frequency sub-bands.}
  \label{fig:ProfileEvolution}
\end{figure*}

\begin{figure}
  \centering
  \includegraphics[width=\columnwidth]{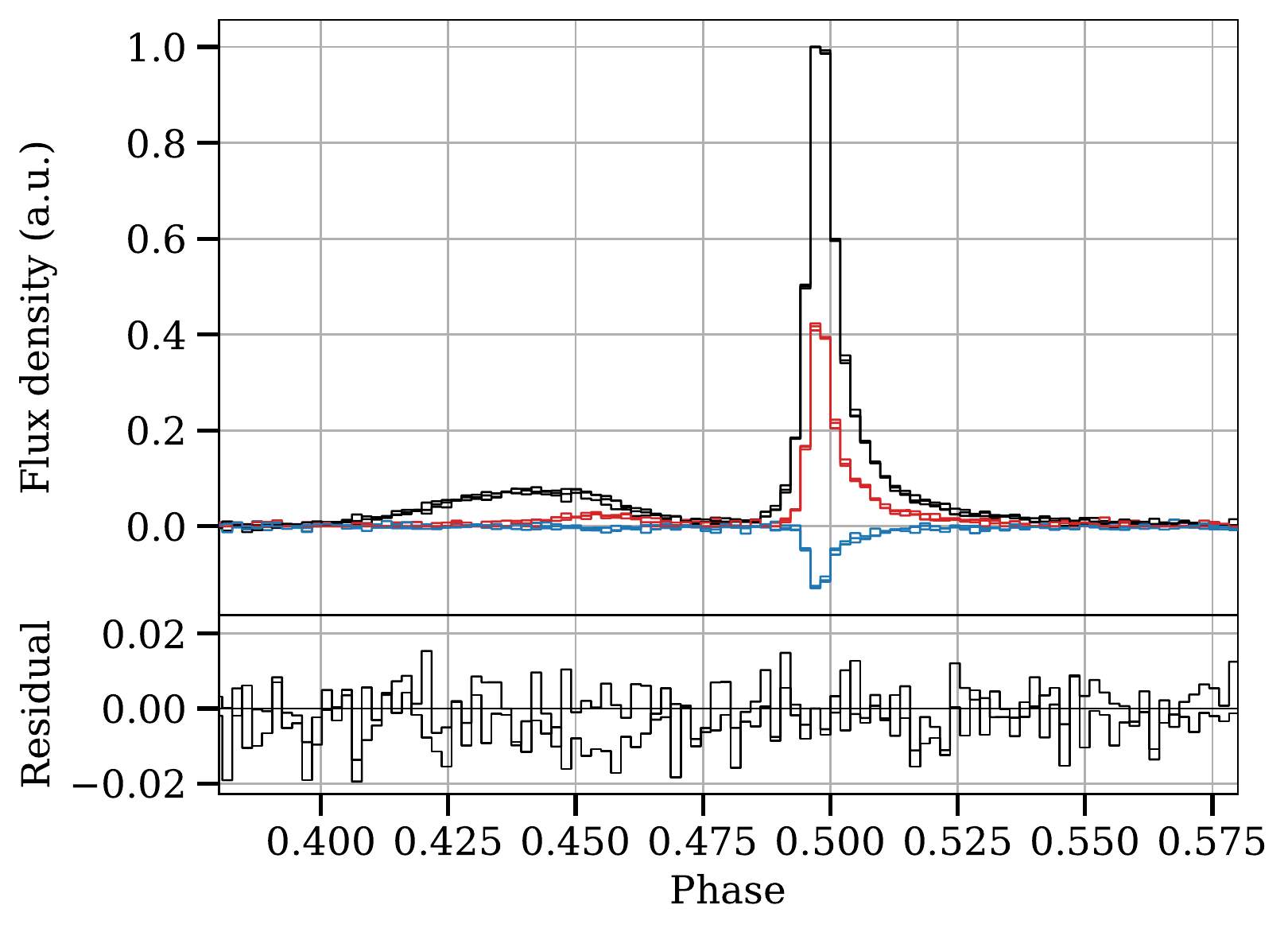}
  \caption{Top panel: Overlay of the normalised band-integrated pulse profiles of PSR~J1452$-$6036 in the combined pre-glitch and the two post-glitch epochs. We zoomed in roughly on the on-pulse phase range and show the total intensity (black), the debiased linear (red) and circular flux densities (blue). The differences are negligible. Bottom panel: Total intensity profile residuals computed by subtracting the pre-glitch profile from the profiles from the two post-glitch epochs. The on-pulse residuals are consistent with being drawn from normal distributions.}
  \label{fig:ProfileDifferences}
\end{figure}

\begin{figure}
  \centering
  \includegraphics[width=\columnwidth]{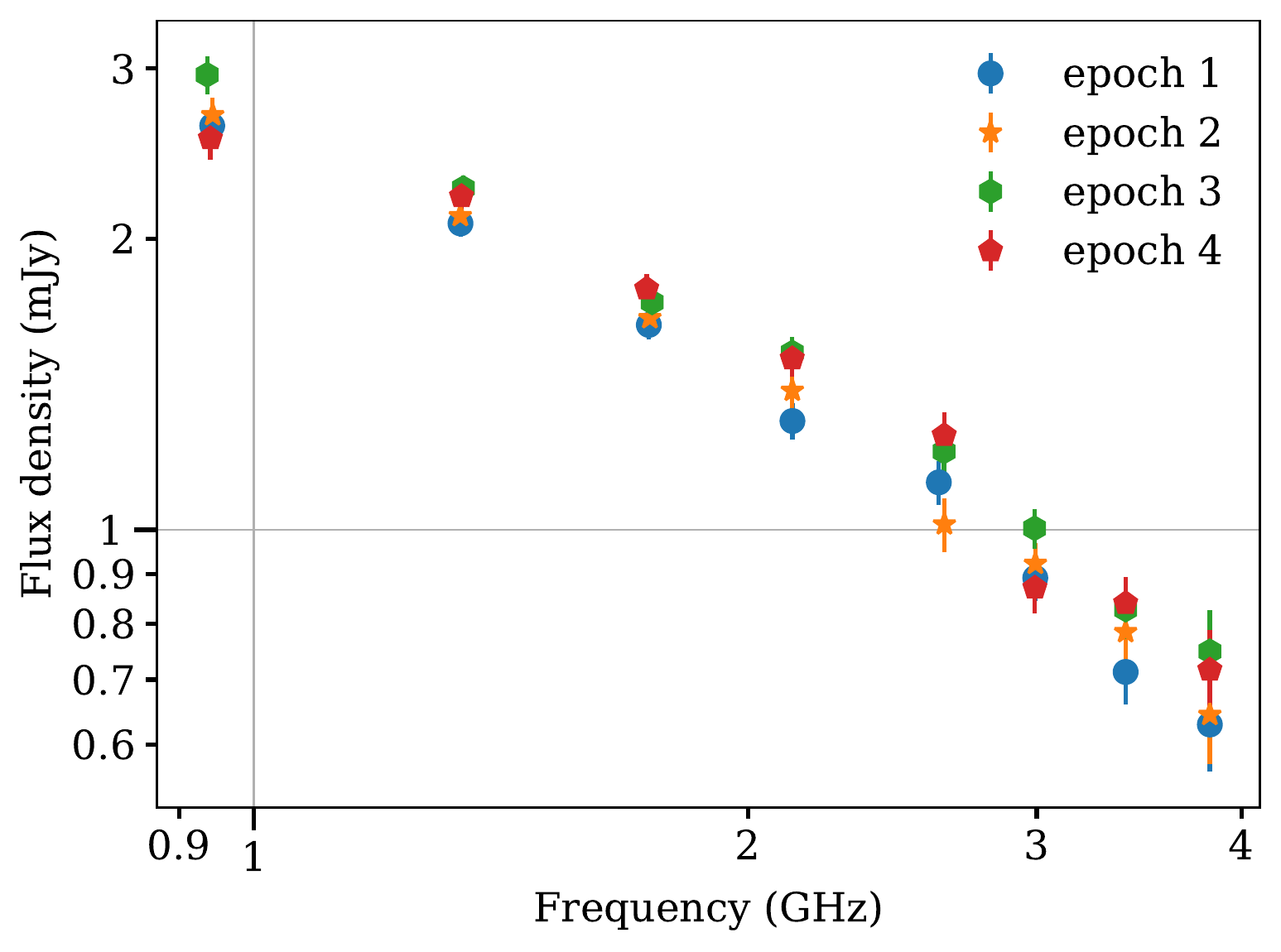}
  \caption{Comparison between the time-integrated spectra in each observing epoch for both profile components, divided into eight frequency sub-bands. The slight overall increase in flux density in epochs 3 and 4 is visible, but the spectral shape stays roughly constant within the uncertainties.}
  \label{fig:SpectrumDifferences}
\end{figure}

The pulsar's profile shows a small amount of evolution across the 3.3-GHz band of the UWL receiver, with the most significant changes visible in the two lowest-frequency sub-bands, where it is dominated by scatter-broadening due to multi-path propagation of the signal in the ISM. We show the combined profile data from all epochs split into eight frequency sub-bands in Fig.~\ref{fig:ProfileEvolution}. The polarimetric profile in the frequency sub-band centred at 1338~GHz agrees well with a reference profile from the literature obtained with the Parkes multi-beam receiver at 1369~MHz \citep{2018Johnston}. However, the value of the polarisation position angle is shifted by about 28~deg in negative direction. The offset in position angle is due to the difference in RM between the literature profile and ours, and the resulting change in Faraday de-rotation of Stokes Q and U; the position angle and RM are covariant. An orthogonal polarisation mode jump between the leading and the main component is apparent in all sub-bands.

To test for any changes in pulse profile morphology, we visually inspected the data in each epoch. Additionally, we subtracted the combined pre-glitch profile from the data obtained at the two post-glitch epochs. As a preparatory step, we aligned the data using the best-fitting ephemeris, subtracted the baselines and normalised the profiles to unit amplitude. We show an overlay of the profiles and their residuals in Fig.~\ref{fig:ProfileDifferences}. A Shapiro-Wilk test for normality \citep{1965Shapiro, 2019Ivezic} in the on-pulse region indicated that we cannot reject the null hypothesis that the residuals have been drawn from a normal distribution, with a p-value of about 0.6 for epochs 3 and 4. The maximum absolute deviations in the on-pulse region are $\sim 0.02$ with RMS values near 0.007 for both epochs. This means that the morphology of the integrated pulse profile stayed unaffected after the glitch down to a precision of $\lesssim 2$~per~cent, or a peak flux density of approximately 2~mJy.

We presented the above profile comparison for the band-integrated data for simplicity, but our conclusions are the same when considering individual frequency sub-bands, although the S/N per sub-band is lower. In Fig.~\ref{fig:SpectrumDifferences} we show a comparison between the absolute flux density spectra in each observing epoch for both profile components and divided into eight frequency sub-bands. The overall slight flux density increase in epochs 3 and 4 is apparent in the sub-banded data too and appears most significant below 1~GHz and above 3~GHz. It happened across the whole bandwidth in epoch 3 and to a slightly lesser extent in epoch 4. The spectral shape stayed roughly constant, as was discussed above.

\section{Discussion}
\label{sec:Discussion}

Our analysis has various caveats. First of all, our data started 3~h after the glitch epoch. While this is short, we were, therefore, unable to detect any radiative changes that decayed completely on time scales faster than that. Second, our data were obtained in pulsar fold-mode and did not resolve individual pulses. In particular, the data from the Medusa backend have 20-s resolution, and the DFB and CASPSR data have 8-s integration times, i.e.\ each integration contains on average about 129 and 52 pulses, respectively. We, therefore, could not constrain possible changes at the single-pulse level. Third, the measured glitch is only of intermediate size. When we compare it with the large glitches of the Vela pulsar that are on average about seven times larger, it could be that only the largest glitches result in the most readily measurable radiative changes, besides our small, but significant, observed flux density increase. In other words, perhaps the glitch was too small to cause any more appreciable changes given the sensitivity of our instrumentation and data. Similarly, two of the four currently published glitches observed in the young, high magnetic-field pulsar J1119$-$6127 even exceeded the ones in the Vela pulsar in fractional size by about a factor of two, with an average size of roughly $5600 \times 10^{-9}$. Interestingly, while anomalous spin-down recoveries were detected in both small and larger glitches in this pulsar, only the large glitch in 2007 resulted in measurable radiative profile changes \citep{2011Weltevrede}, which would support our idea. It would be attractive to re-analyse extant data sets in a similar way that were obtained during or shortly after glitches and are of high S/N to understand whether this is the case. Pulsars that exhibit glitches at the large end of the glitch size distribution (like Vela or PSR~J1119$-$6127) would be the most promising targets if our proposed hypothesis were correct. New and upcoming highly sensitive instruments, such as the MeerKAT telescope array, the Five-hundred-meter Aperture Spherical Telescope (FAST), or the Square-Kilometre Array (SKA), could provide crucial pulse profile data if those facilities were to get on target quickly enough (i.e.\ within a few hours) after glitch events. One could imagine a scheme where small telescopes performed high-cadence monitoring of a sample of glitching pulsars and rapidly triggered the more sensitive facilities in the case of events. Fourth, it could also be that PSR~J1452$-$6036 is not at the right evolutionary stage in the neutron star zoo, where glitches maximally impact radiative parameters. For example, it might show different glitch behaviour if it were more similar to the magnetars or magnetar-like pulsars, as described in the introduction.

\section{Conclusions}
\label{sec:Conclusions}

In this work, we presented new ultra-wideband data of high sensitivity from the Parkes radio telescope that were obtained serendipitously 3~h after an intermediate-sized glitch in the young to middle-aged pulsar PSR~J1452$-$6036. We calibrated the data carefully, and we measured the glitch parameters robustly using MCMC techniques, where we included extant pulsar timing data. We then systematically investigated several pulsar parameters that capture its radiative properties and tested whether they changed significantly after the glitch epoch. We aimed to understand whether the glitch had any impact on the radiative properties of the pulsar. We draw the following conclusions:

\begin{enumerate}
    \item The glitch is only the second reported in this pulsar, after a small glitch in 2009. It has a best-fitting relative size of $\Delta \nu / \nu = 270.52(3) \times 10^{-9}$ and happened on MJD~58600.292(3), i.e.\ 2019-04-27 07:00:28.8.

    \item We measured no significant step in spin-down rate, but derived a $3 \: \sigma$ upper limit of $\left| \Delta \dot{\nu} / \dot{\nu} \right| < 4 \times 10^{-3}$. Similarly, we found no evidence for rapidly-decaying glitch components in our data.
    
    \item Our inferred glitch parameters generally agree with those reported by \citet{2020Lower}, but our Parkes data constrain them better.

    \item The spectral index, the overall spectral shape, the polarisation fractions ($P/I$, $L/I$, and $V/I$), and the RM of the pulsar stayed constant within the uncertainties across the glitch epoch.

    \item The pulse-averaged flux density increased significantly by about 10~per~cent post-glitch in comparison with the pre-glitch values. In the second post-glitch epoch a day later it seemed to have decreased slightly, but was still higher than our pre-glitch measurements. The flux density increase happened across the full bandwidth.

    \item When compared with a sample of reference pulsars observed at the same epochs, the relative change in flux density of PSR~J1452$-$6036 stood out as the one with the highest difference. No comparable trend was seen in any of the reference pulsars. We conclude that the increase is unlikely to be caused by calibration issues.

    \item We compared the increase in flux density with measured long-term modulation indices and theoretical calculations based on ISM properties for the pulsar. We found that diffractive scintillation was well quenched and that only refractive scintillation could explain the change. While we cannot rule out that refractive scintillation was the cause, it seems unlikely, because the increase was near the maximum expected modulation and over a shorter time than what is possible to reconcile with refractive effects.

	\item The morphology of the band-integrated polarimetric pulse profile stayed unaffected after the glitch to a precision of 2~per~cent or better, which corresponds to a peak flux density of $\lesssim 2$~mJy.
\end{enumerate}

\section*{Acknowledgements}

FJ would like to thank Vincent Morello for help with early observations for this project. FJ and BWS acknowledge funding from the European Research Council (ERC) under the  European Union's Horizon 2020 research and innovation programme (grant agreement No. 694745). The Parkes radio telescope is part of the Australia Telescope National Facility which is funded by the Australian Government for operation as a National Facility managed by CSIRO. We acknowledge the Wiradjuri people as the traditional owners of the Observatory site. Part of this work was performed on the OzSTAR national facility at Swinburne University of Technology. OzSTAR is funded by Swinburne and the National Collaborative Research Infrastructure Strategy (NCRIS). We thank the anonymous reviewer and the MNRAS scientific editor Tim Pearson for constructive comments that have improved the paper.

\section*{Data availability}

The data obtained in Parkes project P1011 that are underlying this publication are available from the CSIRO Data Access Portal\footnote{\url{https://data.csiro.au/}}. Reduced data products, such as the best-fitting pulsar ephemeris, are available from our Zenodo repository at \url{https://doi.org/10.5281/zenodo.4593890}. Other data underlying this article will be shared on reasonable request to the corresponding author.



\bibliographystyle{mnras}
\bibliography{glitch_radiative}

\begin{thebibliography}{}
\makeatletter
\relax
\def\mn@urlcharsother{\let\do\@makeother \do\$\do\&\do\#\do\^\do\_\do\%\do\~}
\def\mn@doi{\begingroup\mn@urlcharsother \@ifnextchar [ {\mn@doi@}
  {\mn@doi@[]}}
\def\mn@doi@[#1]#2{\def\@tempa{#1}\ifx\@tempa\@empty \href
  {http://dx.doi.org/#2} {doi:#2}\else \href {http://dx.doi.org/#2} {#1}\fi
  \endgroup}
\def\mn@eprint#1#2{\mn@eprint@#1:#2::\@nil}
\def\mn@eprint@arXiv#1{\href {http://arxiv.org/abs/#1} {{\tt arXiv:#1}}}
\def\mn@eprint@dblp#1{\href {http://dblp.uni-trier.de/rec/bibtex/#1.xml}
  {dblp:#1}}
\def\mn@eprint@#1:#2:#3:#4\@nil{\def\@tempa {#1}\def\@tempb {#2}\def\@tempc
  {#3}\ifx \@tempc \@empty \let \@tempc \@tempb \let \@tempb \@tempa \fi \ifx
  \@tempb \@empty \def\@tempb {arXiv}\fi \@ifundefined
  {mn@eprint@\@tempb}{\@tempb:\@tempc}{\expandafter \expandafter \csname
  mn@eprint@\@tempb\endcsname \expandafter{\@tempc}}}

\bibitem[\protect\citeauthoryear{Anderson \& Itoh}{Anderson \&
  Itoh}{1975}]{1975Anderson}
Anderson P.,  Itoh N.,  1975, \mn@doi [\nat] {10.1038/256025a0}, 256, 25

\bibitem[\protect\citeauthoryear{Andersson, Glampedakis, Ho  \&
  Espinoza}{Andersson et~al.}{2012}]{2012Andersson}
Andersson N.,  Glampedakis K.,  Ho W.,   Espinoza C.,  2012, \mn@doi [Physical
  Review Letters] {10.1103/PhysRevLett.109.241103}, 109, 241103

\bibitem[\protect\citeauthoryear{Archibald, Kaspi, Ng  \& al.}{Archibald
  et~al.}{2013}]{2013Archibald}
Archibald R.,  Kaspi V.,  Ng C.-Y.,   al. E.,  2013, \nat, 497, 591

\bibitem[\protect\citeauthoryear{Bailes et~al.,}{Bailes
  et~al.}{2017}]{2017Bailes}
Bailes M.,  et~al., 2017, \mn@doi [\pasa] {10.1017/pasa.2017.39}, 34, e045

\bibitem[\protect\citeauthoryear{Chukwude \& Urama}{Chukwude \&
  Urama}{2010}]{2010Chukwude}
Chukwude A.,  Urama J.,  2010, \mnras, 406, 1907

\bibitem[\protect\citeauthoryear{Cordes \& Lazio}{Cordes \&
  Lazio}{2002}]{2002Cordes}
Cordes J.~M.,  Lazio T. J.~W.,  2002, preprint (\mn@eprint {arXiv} {0207156})

\bibitem[\protect\citeauthoryear{Dib, Kaspi  \& Gavriil}{Dib
  et~al.}{2008}]{2008Dib}
Dib R.,  Kaspi V.~M.,   Gavriil F.~P.,  2008, \mn@doi [\apj] {10.1086/524653},
  673, 1044

\bibitem[\protect\citeauthoryear{Dodson, McCulloch  \& Lewis}{Dodson
  et~al.}{2002}]{2002Dodson}
Dodson R.,  McCulloch P.,   Lewis D.,  2002, \apjl, 564, L85

\bibitem[\protect\citeauthoryear{{Espinoza}, {Lyne}, {Stappers}  \&
  {Kramer}}{{Espinoza} et~al.}{2011}]{2011Espinoza}
{Espinoza} C.~M.,  {Lyne} A.~G.,  {Stappers} B.~W.,   {Kramer} M.,  2011,
  \mn@doi [\mnras] {10.1111/j.1365-2966.2011.18503.x}, \href
  {https://ui.adsabs.harvard.edu/abs/2011MNRAS.414.1679E} {414, 1679}

\bibitem[\protect\citeauthoryear{Everett \& Weisberg}{Everett \&
  Weisberg}{2001}]{2001Everett}
Everett J. E.~E.,  Weisberg J. M.~W.,  2001, \mn@doi [\apj] {10.1086/320652},
  553, 341

\bibitem[\protect\citeauthoryear{Flanagan}{Flanagan}{1990}]{1990Flanagan}
Flanagan C.,  1990, \mn@doi [\nat] {10.1038/345416a0}, 345, 416

\bibitem[\protect\citeauthoryear{Foreman-Mackey, Hogg, Lang  \&
  Goodman}{Foreman-Mackey et~al.}{2013}]{2013ForemanMackey}
Foreman-Mackey D.,  Hogg D.,  Lang D.,   Goodman J.,  2013, \mn@doi [\pasp]
  {10.1086/670067}, 125, 306

\bibitem[\protect\citeauthoryear{Han, Manchester, Lyne, Qiao  \& van
  Straten}{Han et~al.}{2006}]{2006Han}
Han J.,  Manchester R.,  Lyne A.,  Qiao G.,   van Straten W.,  2006, \mn@doi
  [\apj] {10.1086/501444}, 642, 868

\bibitem[\protect\citeauthoryear{Haskell \& Melatos}{Haskell \&
  Melatos}{2015}]{2015Haskell}
Haskell B.,  Melatos A.,  2015, \mn@doi [International Journal of Modern
  Physics D] {10.1142/S0218271815300086}, 24, 1530008

\bibitem[\protect\citeauthoryear{Hobbs et~al.,}{Hobbs
  et~al.}{2004}]{2004HobbsPMPS}
Hobbs G.,  et~al., 2004, \mn@doi [\mnras] {10.1111/j.1365-2966.2004.08042.x},
  352, 1439

\bibitem[\protect\citeauthoryear{Hobbs, Edwards  \& Manchester}{Hobbs
  et~al.}{2006}]{2006Hobbs}
Hobbs G.,  Edwards R.,   Manchester R.,  2006, \mn@doi [\mnras]
  {10.1111/j.1365-2966.2006.10302.x}, 369, 655

\bibitem[\protect\citeauthoryear{Hobbs et~al.,}{Hobbs et~al.}{2020}]{2020Hobbs}
Hobbs G.,  et~al., 2020, \mn@doi [\pasa] {10.1017/pasa.2020.2}, 37, e012

\bibitem[\protect\citeauthoryear{Hotan, van Straten  \& Manchester}{Hotan
  et~al.}{2004}]{2004Hotan}
Hotan A.,  van Straten W.,   Manchester R.,  2004, \mn@doi [\pasa]
  {10.1071/AS04022}, 21, 302

\bibitem[\protect\citeauthoryear{Ivezi{\'{c}}, Connelly, Vanderplas  \&
  Gray}{Ivezi{\'{c}} et~al.}{2019}]{2019Ivezic}
Ivezi{\'{c}} {\v{Z}}.,  Connelly A.~J.,  Vanderplas J.~T.,   Gray A.,  2019,
  {Statistics, Data Mining, and Machine Learning in Astronomy}.
Princeton University Press

\bibitem[\protect\citeauthoryear{Jankowski, van Straten, Keane, Bailes, Barr,
  Johnston  \& Kerr}{Jankowski et~al.}{2018}]{2018Jankowski}
Jankowski F.,  van Straten W.,  Keane E.,  Bailes M.,  Barr E.,  Johnston S.,
  Kerr M.,  2018, \mn@doi [\mnras] {10.1093/mnras/stx2476}, 473, 4436

\bibitem[\protect\citeauthoryear{Jankowski et~al.,}{Jankowski
  et~al.}{2019}]{2019Jankowski}
Jankowski F.,  et~al., 2019, \mn@doi [\mnras] {10.1093/mnras/sty3390}, 484,
  3691

\bibitem[\protect\citeauthoryear{Johnston \& Kerr}{Johnston \&
  Kerr}{2018}]{2018Johnston}
Johnston S.,  Kerr M.,  2018, \mn@doi [\mnras] {10.1093/mnras/stx3095}, 474,
  4629

\bibitem[\protect\citeauthoryear{{Kaspi}, {Gavriil}, {Woods}, {Jensen},
  {Roberts}  \& {Chakrabarty}}{{Kaspi} et~al.}{2003}]{2003Kaspi}
{Kaspi} V.~M.,  {Gavriil} F.~P.,  {Woods} P.~M.,  {Jensen} J.~B.,  {Roberts}
  M.~S.~E.,   {Chakrabarty} D.,  2003, \mn@doi [\apjl] {10.1086/375683}, \href
  {https://ui.adsabs.harvard.edu/abs/2003ApJ...588L..93K} {588, L93}

\bibitem[\protect\citeauthoryear{Keith, Shannon  \& Johnston}{Keith
  et~al.}{2013}]{2013KeithB}
Keith M.~J.,  Shannon R.~M.,   Johnston S.,  2013, \mn@doi [\mnras]
  {10.1093/mnras/stt660}, 432, 3080

\bibitem[\protect\citeauthoryear{Kramer et~al.,}{Kramer
  et~al.}{2003}]{2003Kramer}
Kramer M.,  et~al., 2003, \mn@doi [\mnras] {10.1046/j.1365-8711.2003.06637.x},
  342, 1299

\bibitem[\protect\citeauthoryear{{Lang}}{{Lang}}{1971}]{1971Lang}
{Lang} K.~R.,  1971, \mn@doi [\apj] {10.1086/150836}, \href
  {https://ui.adsabs.harvard.edu/abs/1971ApJ...164..249L} {164, 249}

\bibitem[\protect\citeauthoryear{Lower et~al.,}{Lower et~al.}{2020}]{2020Lower}
Lower M.,  et~al., 2020, \mn@doi [\mnras] {10.1093/mnras/staa615}

\bibitem[\protect\citeauthoryear{{Luo} et~al.,}{{Luo} et~al.}{2020}]{2021Luo}
{Luo} J.,  et~al., 2020, arXiv e-prints, \href
  {https://ui.adsabs.harvard.edu/abs/2020arXiv201200074L} {p. arXiv:2012.00074}

\bibitem[\protect\citeauthoryear{Lyne, McLaughlin, Keane, Kramer, Espinoza,
  Stappers, Palliyaguru  \& Miller}{Lyne et~al.}{2009}]{2009Lyne}
Lyne A.,  McLaughlin M.,  Keane E.,  Kramer M.,  Espinoza C.,  Stappers B.,
  Palliyaguru N.,   Miller J.,  2009, \mn@doi [\mnras]
  {10.1111/j.1365-2966.2009.15668.x}, 400, 1439

\bibitem[\protect\citeauthoryear{Lyne, Jordan, Graham-Smith, Espinoza, Stappers
   \& Weltevrede}{Lyne et~al.}{2015}]{2015Lyne}
Lyne A.,  Jordan C.,  Graham-Smith F.,  Espinoza C.,  Stappers B.,   Weltevrede
  P.,  2015, \mn@doi [\mnras] {10.1093/mnras/stu2118}, 446, 857

\bibitem[\protect\citeauthoryear{Manchester et~al.,}{Manchester
  et~al.}{2001}]{2001Manchester}
Manchester R.,  et~al., 2001, \mn@doi [\mnras]
  {10.1046/j.1365-8711.2001.04751.x}, 328, 17

\bibitem[\protect\citeauthoryear{Manchester, Hobbs, Teoh  \& al.}{Manchester
  et~al.}{2005}]{2005Manchester}
Manchester R.,  Hobbs G.,  Teoh A.,   al. E.,  2005, \mn@doi [\aj]
  {10.1086/428488}, 129, 1993

\bibitem[\protect\citeauthoryear{McCulloch, Hamilton, McConnell  \&
  al.}{McCulloch et~al.}{1990}]{1990McCulloch}
McCulloch P.,  Hamilton P.,  McConnell  al. E.,  1990, \nat, 346, 822

\bibitem[\protect\citeauthoryear{{Noll}}{{Noll}}{2010}]{2010Noll}
{Noll} C.~E.,  2010, \mn@doi [Advances in Space Research]
  {10.1016/j.asr.2010.01.018}, \href
  {https://ui.adsabs.harvard.edu/abs/2010AdSpR..45.1421N} {45, 1421}

\bibitem[\protect\citeauthoryear{Olausen \& Kaspi}{Olausen \&
  Kaspi}{2014}]{2014Olausen}
Olausen S.,  Kaspi V.,  2014, \mn@doi [\apjs] {10.1088/0067-0049/212/1/6}, 212,
  6

\bibitem[\protect\citeauthoryear{Palfreyman, Dickey, Hotan, Ellingsen  \& van
  Straten}{Palfreyman et~al.}{2018}]{2018Palfreyman}
Palfreyman J.,  Dickey J.,  Hotan A.,  Ellingsen S.,   van Straten W.,  2018,
  \mn@doi [\nat] {10.1038/s41586-018-0001-x}, 556, 219

\bibitem[\protect\citeauthoryear{Petroff, Keith, Johnston, van Straten  \&
  Shannon}{Petroff et~al.}{2013}]{2013Petroff}
Petroff E.,  Keith M.,  Johnston S.,  van Straten W.,   Shannon R.,  2013,
  \mn@doi [\mnras] {10.1093/mnras/stt1401}, 435, 1610

\bibitem[\protect\citeauthoryear{Piekarewicz, Fattoyev  \&
  Horowitz}{Piekarewicz et~al.}{2014}]{2014Piekarewicz}
Piekarewicz J.,  Fattoyev F.,   Horowitz C.,  2014, \mn@doi [\prc]
  {10.1103/PhysRevC.90.015803}, 90, 15803

\bibitem[\protect\citeauthoryear{Radhakrishnan \& Manchester}{Radhakrishnan \&
  Manchester}{1969}]{1969RadhakrishnanB}
Radhakrishnan V.,  Manchester R.~N.,  1969, \mn@doi [Nature]
  {10.1038/222228a0}, 222, 228

\bibitem[\protect\citeauthoryear{Reichley \& Downs}{Reichley \&
  Downs}{1969}]{1969Reichley}
Reichley P.,  Downs G.,  1969, \mn@doi [\nat] {10.1038/222229a0}, 222, 229

\bibitem[\protect\citeauthoryear{{Rickett}, {Coles}  \& {Bourgois}}{{Rickett}
  et~al.}{1984}]{1984Rickett}
{Rickett} B.~J.,  {Coles} W.~A.,   {Bourgois} G.,  1984, \aap, \href
  {https://ui.adsabs.harvard.edu/abs/1984A&A...134..390R} {134, 390}

\bibitem[\protect\citeauthoryear{Ruderman}{Ruderman}{1969}]{1969Ruderman}
Ruderman M.,  1969, \mn@doi [\nat] {10.1038/223597b0}, 223, 597

\bibitem[\protect\citeauthoryear{Sarkissian, Reynolds, Hobbs  \&
  Harvey-Smith}{Sarkissian et~al.}{2017}]{2017Sarkissian}
Sarkissian J.,  Reynolds J.,  Hobbs G.,   Harvey-Smith L.,  2017, \mn@doi
  [\pasa] {10.1017/pasa.2017.19}, 34, e027

\bibitem[\protect\citeauthoryear{{Sarkissian}, {Hobbs}, {Reynolds},
  {Palfreyman}  \& {Olney}}{{Sarkissian} et~al.}{2019}]{2019Sarkissian}
{Sarkissian} J.,  {Hobbs} G.,  {Reynolds} J.,  {Palfreyman} J.,   {Olney} S.,
  2019, The Astronomer's Telegram, \href
  {https://ui.adsabs.harvard.edu/abs/2019ATel12466....1S} {12466, 1}

\bibitem[\protect\citeauthoryear{Shapiro \& Wilk}{Shapiro \&
  Wilk}{1965}]{1965Shapiro}
Shapiro S.,  Wilk M.~B.,  1965, \mn@doi [Biometrika]
  {10.1093/biomet/52.3-4.591}, 52, 591

\bibitem[\protect\citeauthoryear{Sieber}{Sieber}{1982}]{1982Sieber}
Sieber W.,  1982, \aap, 113, 311

\bibitem[\protect\citeauthoryear{Sotomayor-Beltran et~al.,}{Sotomayor-Beltran
  et~al.}{2013}]{2013Sotomayor}
Sotomayor-Beltran C.,  et~al., 2013, \mn@doi [Astronomy and Astrophysics]
  {10.1051/0004-6361/201220728}, 552, 1

\bibitem[\protect\citeauthoryear{Stinebring, Smirnova, Hankins, Hovis, Kaspi,
  Kempner, Myers  \& Nice}{Stinebring et~al.}{2000}]{2000Stinebring}
Stinebring D.,  Smirnova T.,  Hankins T.,  Hovis J.,  Kaspi V.,  Kempner J.,
  Myers E.,   Nice D.,  2000, \mn@doi [\apj] {10.1086/309201}, 539, 300

\bibitem[\protect\citeauthoryear{{Th{\'e}bault} et~al.,}{{Th{\'e}bault}
  et~al.}{2015}]{2015Thebault}
{Th{\'e}bault} E.,  et~al., 2015, \mn@doi [Earth, Planets, and Space]
  {10.1186/s40623-015-0228-9}, \href
  {https://ui.adsabs.harvard.edu/abs/2015EP&S...67...79T} {67, 79}

\bibitem[\protect\citeauthoryear{Weltevrede, Johnston  \& Espinoza}{Weltevrede
  et~al.}{2011}]{2011Weltevrede}
Weltevrede P.,  Johnston S.,   Espinoza C.~M.,  2011, \mn@doi [Monthly Notices
  of the Royal Astronomical Society] {10.1111/j.1365-2966.2010.17821.x}, 411,
  1917

\bibitem[\protect\citeauthoryear{{Yu} et~al.,}{{Yu} et~al.}{2013}]{2013Yu}
{Yu} M.,  et~al., 2013, \mn@doi [\mnras] {10.1093/mnras/sts366}, \href
  {https://ui.adsabs.harvard.edu/abs/2013MNRAS.429..688Y} {429, 688}

\makeatother
\end{thebibliography}



\appendix



\bsp	
\label{lastpage}
\end{document}